\documentclass{elsart}

    \usepackage{latexsym,bm,amsmath,amssymb,amsfonts}
    \usepackage{epsfig,graphics,graphicx}
    \usepackage{makeidx}
    \usepackage{citesort}

    \newcommand{\abar}{\bar{\alpha}_s}
    \newcommand{\del}{\partial}

    \newcommand{\wt}[1]{\widetilde{#1}}

\long\def\comment#1{ }

\newcommand{\nn}{\nonumber\\ }
\newcommand{\beq}{\begin{eqnarray}}
\newcommand{\eeq}{\end{eqnarray}}
\newcommand{\be}{\vspace{-.4cm}\begin{eqnarray}}
\newcommand{\ee}{\vspace{-.5cm}\end{eqnarray}}

\newcommand{\lan}{\langle}
\newcommand{\ran}{\rangle}

\def\del{\partial}

\newcommand{\cal}{\mathcal} 

\newcommand{\tr}{{\rm tr}}
\newcommand{\Tr}{{\rm Tr}}

\def\simge{\mathrel{%
   \rlap{\raise 0.511ex \hbox{$>$}}{\lower 0.511ex \hbox{$\sim$}}}}
\def\simle{\mathrel{
   \rlap{\raise 0.511ex \hbox{$<$}}{\lower 0.511ex \hbox{$\sim$}}}}

\newcommand{\x}{\bm x}
\newcommand{\y}{\bm y}

\newcommand{\z}{\bm z}

\begin{document}

\begin{flushright}
~\vspace{-1.25cm}\\
{\small\sf SACLAY--T05/089
\\  BNL-NT-05/14}

\end{flushright}
\vspace{2.cm}

\begin{frontmatter}

\parbox[]{16.0cm}{ \begin{center}
\title{Color dipoles from Bremsstrahlung \\
in QCD evolution at high energy}

\author{Y.~Hatta$^{\rm a}$},
\author{E.~Iancu$^{\rm b,}$\thanksref{th2}},
\author{L.~McLerran$^{\rm a,c}$},
\author{A.~Stasto$^{\rm c,d}$}

\address{$^{\rm a}$ RIKEN BNL Research Center, Brookhaven National Laboratory,
Upton, NY 11973, USA}

\address{$^{\rm b}$Service de Physique Theorique, Saclay,
F-91191 Gif-sur-Yvette, France}

\address{$^{\rm c}$ Physics Department, Brookhaven
National Laboratory, Upton, NY 11973, USA}

\address{$^{\rm d}$ H. Niewodnicza\'nski Institute of Nuclear Physics, \\
ul. Radzikowskiego 152, 31-342 Krak\'ow Poland}

\thanks[th2]{Membre du Centre National de la Recherche Scientifique
(CNRS), France.}

\date{\today}
\vspace{0.8cm}
\begin{abstract}
We show that the recently developed Hamiltonian theory for high
energy evolution in QCD in the dilute regime and in the presence
of Bremsstrahlung is consistent with the color dipole picture in
the limit where the number of colors $N_c$ is large. The color
dipoles are quark--antiquark pairs which can radiate arbitrarily
many soft gluons, and the evolution consists in the splitting of
any such a dipole into two. We construct the color glass weight
function of an onium as a superposition of color dipoles, each
represented by a pair of Wilson lines. We show that the action of
the Bremsstrahlung Hamiltonian on this weight function and in the
large--$N_c$ limit generates the evolution expected from the
dipole picture. We construct the dipole number operator in the
Hamiltonian theory and deduce the evolution equations for the
dipole densities, which are again consistent with the dipole
picture. We argue that the Bremsstrahlung effects beyond two gluon
emission per dipole are irrelevant for the calculation of
scattering amplitudes at high energy.

\end{abstract}
\end{center}}

\end{frontmatter}
\newpage

\section{Introduction}
\setcounter{equation}{0} \label{S_Intro}

Our purpose in this paper is to demonstrate that the perturbative
evolution of a {\em dilute} hadronic system with increasing energy
can be effectively described as the evolution of a system of {\em
color dipoles} provided we consider the limit in which the number
of colors $N_c$ is large and that the energy remains low enough
for the evolved system to be still dilute. By `dilute' we mean
that the hadronic system is {\em non--saturated} : the gluon
density is low enough for the recombination processes to be
unimportant.

At an abstract, {\em wavefunction}, level, where the evolution is
viewed in terms of the partonic content of the system and not of
its interactions, the effectiveness of the dipole picture has been
of course demonstrated in the original paper by Mueller
\cite{AM94}. Here, however, we are primarily interested in the
problem of {\em scattering} --- the dilute hadronic system is the
{\em target} which scatters with an external {\em projectile},
itself assumed to be dilute --- and to that purpose we need to
specify not only the evolution of the partons in the target, but
also the way how an individual parton (gluon or dipole) couples to
the projectile. In previous applications of the dipole picture to
scattering \cite{AM95,IM031,IT04,MSW05,IT05,BIIT05}, one has
always assumed that a target dipole which partakes in the
collision exchanges exactly two gluons --- i.e., it undergoes {\em
single scattering} --- with the projectile. Whereas this
approximation is indeed justified at relatively high energy, where
the large dipole density in the target favors the multiple
scattering with {\em different} dipoles \cite{AM95} (see also the
discussion in Sect. 5 below), this is not really correct at low
energies, where e.g. the double scattering off a same target
dipole via four gluon exchange competes with the scattering off
two different dipoles, via twice two gluon exchange.

Notice that multiple gluon exchanges between a target dipole and
the projectile correspond to many gluon radiation from the quark
and the antiquark legs of the dipole, that is, to {\em gluon
Bremsstrahlung}. The theoretical description of the Bremsstrahlung
in the high--energy evolution of a dilute QCD system has recently
became available \cite{KL05,KL3,BREM}, but its relation with the
dipole picture at large $N_c$ has not been established so far, in
spite of previous attempts \cite{KL4} which emphasized the
complexity of the problem. It is our present objective to fully
clarify this problem by showing that, at large $N_c$, the
evolution generated by the Bremsstrahlung Hamiltonian in Refs.
\cite{KL05,KL3,BREM} can be recast in the language of the dipole
picture.

Specifically, we shall find that, for large $N_c$, the
wavefunction of the dilute target (the `onium') can be described
as a collection of color dipoles, where each dipole is allowed to
radiate arbitrarily many small--$x$ gluons in the eikonal
approximation. At a mathematical level, a dipole is represented by
a pair of Wilson lines (one for the quark and the other one for
the antiquark) built with the color field of the radiated gluons.
The evolution of the system with increasing energy is governed by
the action of the Bremsstrahlung Hamiltonian on the `onium'
wavefunction. As we shall see, at large $N_c$ this evolution
proceeds through dipole splitting: one pair of Wilson lines splits
into two such pairs which have one common transverse coordinate
(at the position of the emitted gluon).

The formulation of the dipole picture that we shall naturally
arrive at is of the {\em color glass} type
\cite{MV,RGE,CGCreviews} : The dilute target is effectively
represented as a stochastic ensemble of classical color fields ---
the fields radiated by the dipoles
--- which are distributed according to a functional
`weight function' whose evolution we shall compute. This evolution
can be reformulated in terms of probabilities for the dipole
configurations (or, equivalently, in terms of dipole $n$--body
densities, with $n\ge 1$), and then it reduces to the original
wavefunction evolution\footnote{More precisely, the formulation of
the `onium' wavefunction evolution which will naturally emerge
from our calculations is that  in Refs.
\cite{IM031,LL03,LL04,IT04}.} by Mueller \cite{AM94}, as expected.
But the color glass formulation turns out to be more convenient
for applications to the scattering problem, as it gives directly
the distribution of the target color fields to which couple the
projectile.

At this point we should remind that a color glass formulation of
the dipole picture has been already given in Refs.
\cite{IM031,MSW05}, but under the assumption that each dipole can
radiate only two gluons. When performing the corresponding
approximation on our subsequent results (this amounts to expanding
the Wilson lines to lowest non--trivial order), we shall recover
the weight function of Refs. \cite{IM031,MSW05}, as expected.
However, this does not mean that the Bremsstrahlung Hamiltonian
$H_{\rm BREM}$
should mechanically reduce at large $N_c$ to the `Dipole Model'
Hamiltonian $H_{\rm MSW}$ introduced by Mueller, Shoshi and Wong
\cite{MSW05}. Rather, these two Hamiltonians describe the same
physical process
--- the splitting of one dipole into two --- but by acting on {\em
different} dipole operators: $H_{\rm MSW}$ acts in the Hilbert
space of {\em bare} dipoles, by which we mean the dipoles which
are allowed to radiate only two gluons, whereas $H_{\rm BREM}$
acts on dipoles which are fully {\em dressed} by the radiation. In
fact, the ``large--$N_c$ limit of $H_{\rm BREM}$'' is not a
well--defined concept by itself: The simplifications appropriate
at large $N_c$ can be performed only in the process of acting with
$H_{\rm BREM}$ on the onium weight function, or on
gauge--invariant correlations expressing dipole densities.

Since both $H_{\rm BREM}$ and $H_{\rm MSW}$ describe dipole
splitting (at large $N_c$), they are equivalent in so far as the
evolution of the onium {\em wavefunction} is concerned: they
generate the same evolution equations for the dipole densities (or
the associated probabilities), which moreover coincide with the
corresponding equations previously derived within the `abstract'
dipole picture \cite{AM94,AM95,IM031,LL04,IT04}. However, since
the physical information encoded in the dipole operator is
different in the two theories --- this is richer for $H_{\rm
BREM}$, where a single dipole is allowed to radiate arbitrarily
many gluons ---, and moreover this information is relevant for the
scattering, it follows that the evolution equations for {\em
scattering amplitudes} generated by the two Hamiltonians will be
different in general. For instance, the scattering between the
onium and two external dipoles may proceed either via twice two
gluon exchange with two different internal dipoles, or via four
gluon exchange with a single internal dipole. $H_{\rm MSW}$
describes only the former process, and thus generates the
`fluctuation' terms in the evolution equations with Pomeron loops
\cite{IT04,IT05}, while $H_{\rm BREM}$ describes both of them, and
thus in principle it generates more complete equations. Still, as
mentioned before, and will be demonstrated in Sect. 5 below, the
latter process is suppressed at high energy with respect to the
former one, and thus is irrelevant for the study of the approach
towards saturation and the unitarity limit. A similar conclusion
has been recently reached by Marquet, Mueller, Shoshi and Wong
\cite{AMprivate,MMSW05}.

The relation between $H_{\rm BREM}$ and $H_{\rm MSW}$ alluded to
above is in fact similar to that between the JIMWLK Hamiltonian
and its `Pomeron' approximation at large $N_c$, as discussed in
Ref. \cite{BIIT05}. The JIMWLK Hamiltonian \cite{JKLW97,RGE,W}
describes non--linear effects (gluon recombination, or merging) in
the evolution of the target wavefunction towards saturation.
Alternatively, when acting on scattering operators built with
Wilson lines, it describes gluon splitting in the projectile
followed by the scattering between the products of this splitting
and the target. At large $N_c$, it is convenient to consider a
projectile which is itself made with dipoles. Then, as shown in
Ref. \cite{BIIT05}, the action of $H_{\rm JIMWLK}$  on a {\em
dressed} external dipole (where the `dressing' now refers to
multiple scattering, as encoded in the Wilson lines) is equivalent
--- in the sense of generating the same evolution equations for
the scattering amplitudes at large $N_c$ --- to that of a simpler
`Pomeron merging' Hamiltonian on a {\em bare} external dipole,
i.e., a dipole which exchanges only two gluons with the target.
This close correspondence between the two problems
--- the JIMWLK evolution in the high density regime and the
evolution including Bremsstrahlung in the dilute regime --- should
not came as a surprise, in view of the {\em duality} between
$H_{\rm BREM}$ and $H_{\rm JIMWLK}$ on one side \cite{KL3,BREM},
and between $H_{\rm MSW}$ and the Pomeron--merging Hamiltonian on
the other side \cite{BIIT05}. Our subsequent analysis will shed
more light on this correspondence, by showing that the duality
holds, more precisely, between the action of $H_{\rm JIMWLK}$ on
{\em scattering} operators for {\em external} dipoles and,
respectively, that of $H_{\rm BREM}$ on {\em radiation} operators
for the {\em internal} dipoles. In brief, the action of $H_{\rm
JIMWLK}$ on the projectile is dual to that of $H_{\rm BREM}$ on
the target.

The plan of the paper is as follows: In Sect. 2 we briefly review
the Hamiltonian formulation of the high--energy evolution in the
dilute regime and in the presence of Bremsstrahlung, then explain
the action of $H_{\rm BREM}$ in terms of Poisson brackets (or
`commutators'), and finally demonstrate as a simple exercise that
the action of this Hamiltonian on the 2--point function of the
color charge density generates the BFKL equation, as expected.
Sect. 3 contains our main results, namely the construction of the
onium weight function in terms of dressed dipoles, and the proof
of the fact that the action of $H_{\rm BREM}$ on this weight
function generates the dipole evolution at large $N_c$. In Sect.
4, we show how to express dipole number densities in terms of
gauge--invariant correlations of the color charge. We introduce
the dipole number operator, and verify that, under the action of
$H_{\rm BREM}$ (at large $N_c$), the dipole densities obey the
evolution equations expected in the dipole picture. In Sect. 5 we
consider the onium scattering with two external dipoles and show
that the elementary processes which involve the double scattering
of a same target dipole are subdominant at high energy. Finally,
Sect. 6 contains our conclusions.

\section{QCD evolution in the low density regime: Bremsstrahlung Hamiltonian}
\setcounter{equation}{0} \label{S_BREM}

The physical problem that we have in mind is that of the
scattering between a dilute hadronic system propagating in the
positive $z$ (or positive $x^+$) direction --- the right--moving
target --- and a system of color dipoles moving in the negative
$z$ (or positive $x^-$) direction --- the left--moving projectile.
When the projectile is made of a single dipole, the scattering
measures the gluon distribution in the target, i.e., the average
gluon number density. A projectile made with several dipoles will
also probe {\em fluctuations} in the gluon number. In the limit
where the number of colors $N_c$ is large, we shall be able to
effectively describe the gluon distribution in the target in
terms of (internal) dipoles. Then the scattering with the external
system of dipoles will measure the {\em dipole distribution} in
the target, that is, the average dipole number density and the
corresponding fluctuations.

The operator expressing the scattering amplitude for an external
dipole $(\x,\y)$ in the eikonal approximation reads:
 \be\label{Tdipole} T(\x,\y) \,=\,1\,-\,
\frac{1}{N_c}\,\tr \big(V^\dag({\x}) V({\y})\big)\, \simeq \,
  \frac{g^2}{4N_c}\,
 \big(\alpha^a({\x})-\alpha^a({\y})\big)^2,\ee
where the second, approximate, equality holds in the case where
the target is dilute, which is the case of interest for us here.
In this equation, the Wilson line
 \be \label{up}
  V^\dag(\x)={\mbox P}\exp \left\{
ig\int dx^- \alpha_a(x^-,\x)\,t^a \right\}\,,
 \ee
(the $t^a$'s are the generators of the SU$(N_c)$ algebra in the
fundamental representation and the symbol P denotes path--ordering
in $x^-$) describes the eikonal scattering between a quark with
transverse coordinate $\x$ and the light--cone component
$A^+_a\equiv \alpha_a(x^-,\x)$ of the color field in the target.
Similarly, $V({\y})$ describes the scattering of the antiquark in
the dipole. Furthermore,
 \be\label{alphaT} \alpha_a(\x)\,\equiv\,\int
 dx^-\,\alpha_a(x^-,\x)\,\ee 
is the effective color field in the transverse plane, as obtained
after integrating over the longitudinal profile of the target, and
is related to the corresponding color charge density in the target
$\rho_a(\x)$ via the two--dimensional Poisson equation
  \be\label{Poisson} -
\nabla^2_\perp \alpha_a(\x)
 \,=\,\rho_a(\x) \,.\ee
Note that the time variable $x^+$ is suppressed in the equations
above, since the scattering is quasi--instantaneous; thus,
$\alpha$ and $\rho$ are the color field and charge at the time of
scattering, which is $x^+=\infty$ in the conventions of Ref.
\cite{BREM} : $\alpha \equiv\alpha_\infty$ and $\rho\equiv
\rho_\infty$.

The physical scattering amplitude for a projectile made with $k$
dipoles is obtained as $\langle T^{(k)}\rangle_\tau = \langle
T(\x_1,\y_1)T(\x_2,\y_2)\cdots T(\x_k,\y_k)\rangle_\tau$, where
$\tau\sim\ln s$ is the rapidity gap between the projectile and the
target (note that we use a Lorentz frame in which most of the
total energy is carried by the target) and the brackets denote the
average over the target wavefunction, which in the spirit of the
color glass formalism \cite{CGCreviews} is computed as an average
over the color charge $\rho$ with weight function $Z_\tau[\rho]$.
E.g., for a single dipole:
 \be\label{AVTdipole} \langle T(\x,\y)\rangle_\tau\,\simeq
 \,\int D[\rho]\,\,\frac{g^2}{4N_c}\,
 \big(\alpha^a({\x})-\alpha^a({\y})\big)^2\,
 Z_{\tau}[\rho]\,.\ee
To account for the correlations induced by Bremsstrahlung, the
color charge density should be treated as a time--dependent
variable \cite{KL05,KL3,BREM}, so $Z_{\tau}[\rho]$ is a functional
of $\rho_a(x^+,\x)$ (see below for more details).

According to the above equations, in order to compute physical
scattering amplitudes and their evolution with $\tau$, it is
sufficient to know the target weight function $Z_{\tau}[\rho]$ and
the corresponding evolution equation. The latter can be compactly
written as
 \be\label{RGEZ}
 \frac{\partial}{\partial \tau} \,Z_{\tau}[\rho]\,=\,-\,
 H_{\rm BREM}\,Z_{\tau}[\rho]\,,\ee
with the following Hamiltonian for color glass evolution in the
dilute regime and in the presence of Bremsstrahlung
\cite{KL05,KL3,BREM} :
 \begin{align} \label{HBREM}
    H_{\rm BREM}
    = \frac{1}{(2\pi)^3}\,
    \int\limits_{\bm{x}\bm{y}\bm{z}}
    \mathcal{K}_{\bm{x}\bm{y}\bm{z}}\,
    \rho^a_{\infty}(\bm{x})
    \left[1 +
    \wt W_{\bm{x}}
    \wt W^{\dagger}_{\bm{y}}-
    \wt W_{\bm{x}}
    \wt  W^{\dagger}_{\bm{z}}-
    \wt W_{\bm{z}}
    \wt W^{\dagger}_{\bm{y}} \right]^{ab}
    \rho^b_{\infty}(\bm{y}).
\end{align}
(For more clarity, we have temporarily restored the subscript
$\infty$ denoting the $x^+$ variable of the field $\rho$.) In this
equation,
  \be\label{Kdef} {\mathcal
K}({\bm{x}, \bm{y}, \bm{z}}) \,\equiv \,
   \frac{(\bm{x}-\bm{z})\cdot(\bm{y}-\bm{z})}{
     (\bm{x}-\bm{z})^2 (\bm{z}-\bm{y})^2}\,.\ee
Furthermore, $\wt W$ and $\wt W^{\dagger}$ are all--order
differential operators defined as, e.g.,
 \begin{align}\label{Wdef}
\wt W(\bm{x}) \; = \; {\rm T} \,\exp\left\{-g \int
 dx^+\, T^a\,\frac{\delta}{\delta \rho_a(x^+,\bm{x})}  \right\}
\end{align}
where T denotes time--ordering and the color matrices $T^a$ are in
the adjoint representation.
Physically,  $\wt W$ and $\wt W^{\dagger}$ are Wilson lines
describing the radiation of an arbitrary number of small--$x$
gluons in the eikonal approximation from a single gluon with a
larger value of $x$. (As usual, $x$ denotes the longitudinal
momentum fraction carried by a gluon, with $\tau=\ln 1/x\,$; see
Ref. \cite{BREM} for details.)  When Eq.~(\ref{Wdef}) is expanded
in powers of $\delta/\delta\rho$, each such a power describes the
creation of a source for the emission of one gluon (cf.
Eq.~(\ref{Poisson})). Since these sources are dilute, the radiated
fields are never strong. However, by keeping terms of all orders
in $\delta/{\delta \rho}$ in the Wilson lines, and thus in the
Hamiltonian (\ref{HBREM}), one includes in the evolution processes
like $2 \to n$ gluon splitting through which the $n$--point
functions of $\rho$ with $n>2$ get built from the 2--point
function $\lan\rho\rho\ran$ in the dilute regime.

For what follows, the explicit representation (\ref{Wdef}) for
$\wt W$ is not so important. Rather, what matters is the structure
of the {\em Poisson brackets} which define the action of the
Hamiltonian (\ref{HBREM}) on the Hilbert space of the
operators\footnote{In what follows, we shall use the notation $W$
for the temporal Wilson line in a generic, or in the fundamental,
representations, and we shall keep the more specific notation $\wt
W$ for the adjoint representation.} ${\cal O}[\rho_\infty,W]$:
 \be\label{PB}
 \big[\rho^a_\infty(\bm{x}), \,
 \wt W_{bc}(\bm{y})\big]&\,=\,&g(T^a\wt W(\bm{x}))_{bc}\,
 \delta^{(2)}({\bm{x-y}}),
\nn\big [\rho^a_\infty(\x),\, \rho_\infty^b(\y) \big]
  &\,=\,& - igf^{abc}\rho_\infty^c(\x) \,\delta^{(2)}(\x-\y),\nn
 \big [\wt W_{ab}(\bm{x}), \wt W_{cd}(\bm{y})\big]&\,=\,&0\,.\ee
The first two equations above show that the color charges
$\rho^a_{\infty}$ act as infinitesimal gauge rotations of the
Wilson line $W$ at its end points, that is, as Lie derivatives.
One can check that these commutation relations satisfy the right
properties expected for Poisson brackets, in particular, they obey
the Jacobi identity.

The non--commutativity of the color charge variables
$\rho^a_\infty(\bm{x})$ with themselves is a source of potential
difficulties (like ambiguities in the ordering of the operators)
in the construction of the correlation functions and of the color
glass weight function \cite{KL05,KL3,KL4,BREM}. Still, as we shall
see, this problem can be systematically avoided in the
large--$N_c$ limit, where we shall be able to explicitly construct
the weight function $Z_\tau[\rho]$ in a way which is free of
ambiguities.

The formal evolution equation for some operator ${\cal
O}[\rho_\infty,W]$ is defined by its Poisson bracket with $H_{\rm
BREM}$ :
 \be\label{EVOLO}
 \frac{\partial}{\partial \tau} \, {\cal
 O}[\rho_\infty,W]\,=\,\big[
 H_{\rm BREM}\,,{\cal O}\big]\,.\ee
Note that for a given ordering of the operators which compose
${\cal O}$, the operation above is unambiguous. However, its
result will generally change when we permute the operators within
the definition of ${\cal O}$. Once again, this difficulty will not
show up at large $N_c$.

One can check that, for physical observables at least,
Eq.~(\ref{EVOLO}) is indeed consistent with the equation
(\ref{RGEZ}) for the evolution of the  weight function together
with the definition (\ref{AVTdipole}) of the color glass average.
To see this, note first that the physical observables are
gauge--invariant, which in the present context means that they are
invariant under the gauge transformations dependent upon $x^+$
\cite{BREM}. When the action of $H_{\rm BREM}$ is restricted to
such gauge--invariant operators, Eq.~(\ref{HBREM}) can be
equivalently replaced by \cite{ODDERON}
\begin{align}\label{HMBREM}
    H_{\rm BREM}
    = \frac{-1}{16\pi^3}\,
    \int\limits_{\bm{x}\bm{y}\bm{z}}
    \mathcal{M}_{\bm{x}\bm{y}\bm{z}}\,
    \left[1 +
    \wt W_{\bm{x}}
    \wt W^{\dagger}_{\bm{y}}-
    \wt W_{\bm{x}}
    \wt W^{\dagger}_{\bm{z}}-
    \wt W_{\bm{z}}
    \wt W^{\dagger}_{\bm{y}} \right]^{ab}
   \rho^a_{\infty}(\bm{x}) \rho^b_{\infty}(\bm{y}),
\end{align}
with $\mathcal{M}_{\bm{x}\bm{y}\bm{z}}$ denoting the dipole kernel
\cite{AM94} :
 \be \label{Mdef} {\mathcal M}({\bm{x}},{\bm{y}},{\bm
z})\,\equiv\, \frac{(\bm{x}-\bm{y})^2}{(\bm{x}-\bm{z})^2
(\bm{z}-\bm{y})^2}\,= {\mathcal K}_{\bm{x} \bm{x} \bm{z}} +
{\mathcal
 K}_{\bm{y} \bm{y} \bm{z}} - 2{\mathcal K}_{\bm{x} \bm{y} \bm{z}} \,.\ee
(This can be proven through the `dual' version of the arguments
used in Ref. \cite{ODDERON} for the case of the JIMWLK
Hamiltonian.) From Eq.~(\ref{PB}) one can check that, in the
presence of the dipole kernel, the color charge operators in
Eq.~(\ref{HMBREM}) can be freely commuted through the Wilson lines
there; in writing Eq.~(\ref{HMBREM}) we have used this freedom to
commute both factors of $\rho$ fully to the right, which is
convenient for the subsequent manipulations. Also, the
non--commutativity between $\rho^a_{\infty}(\bm{x})$ and $
\rho^b_{\infty}(\bm{y})$ plays no role in Eq.~(\ref{HMBREM}) since
the Wilson--line part of the integrand,
 \be\label{defh}
 h_{ab}(\x,\y,\z) \equiv \left[1 +
    \wt W_{\bm{x}}
    \wt W^{\dagger}_{\bm{y}}-
    \wt W_{\bm{x}}
    \wt  W^{\dagger}_{\bm{z}}-
    \wt W_{\bm{z}}
    \wt W^{\dagger}_{\bm{y}} \right]^{ab},\ee
is symmetric under the simultaneous exchange $a \leftrightarrow b$
and $\x \leftrightarrow \y$.

Note furthermore that, if the operator ${\cal O}[\rho_\infty,W]$
contains a factor of $W$ on the left of all the other operators,
then when computing the color glass average $\lan {\cal
O}[\rho_\infty,W]\ran_\tau$ this factor can be replaced by one.
Indeed, when expanding the exponential in Eq.~(\ref{Wdef}), all
the terms but the first one yield total derivatives which vanish
after integration over $\rho$. In the subsequent manipulations, it
will be often convenient to `normal--order' the operators by
pushing factors of $W$ all the way to the left and the replacing
them by one when computing the average.

We are now prepared to check that Eqs.~(\ref{RGEZ}) and
(\ref{EVOLO}) are consistent with each other. To that aim, take
the color glass average in Eq.~(\ref{EVOLO}). This involves
 \be
 &{}& \int D[\rho]\,\,\big[h_{ab}(\x,\y,\z)\rho^a_{\infty}(\bm{x})
 \rho^b_{\infty}(\bm{y}),\, {\cal O}\big]\,Z_{\tau}[\rho]\nn
 &{}& =\int D[\rho]\,\,\big(h_{ab}\rho^a_{\infty}
 \rho^b_{\infty}{\cal O}\,-\,{\cal O}h_{ab}\rho^a_{\infty}
 \rho^b_{\infty}\big)\,Z_{\tau}[\rho]\nn
 &{}& =\int D[\rho]\,\,\big(-\,{\cal O}h_{ab}\rho^a_{\infty}
 \rho^b_{\infty}\big)\,Z_{\tau}[\rho]\,\longrightarrow
 \int D[\rho]\,\,{\cal O}\,\,(-\,H_{\rm BREM}Z_{\tau}[\rho])
 \,,\ee
where we have used the fact that $h[W]\to 0$ when $W\to 1$. By the
same argument, we deduce that the average of Eq.~(\ref{EVOLO})
involves only the commutator $[h, {\cal O}]$ :
 \be\label{AVEVOLO}
 \frac{\partial}{\partial \tau} \, \lan {\cal
 O}\ran_\tau\,=\, \frac{-1}{16\pi^3}\,
   \int D[\rho]\, \int\limits_{\bm{x}\bm{y}\bm{z}}
    \mathcal{M}_{\bm{x}\bm{y}\bm{z}}\,
    \,\big[h_{ab}(\x,\y,\z)
    \,,{\cal O}\big]\,\rho^a_{\infty}(\bm{x})
     \rho^b_{\infty}(\bm{y})\,Z_{\tau}[\rho].\ee

As a first, relatively simple, application of the above formalism,
let us derive in this way the BFKL equation. We shall argue later
that, at large $N_c$ and for $\x\ne \y$, the charge--charge
correlator $\lan \rho^a(\bm{x}) \rho^a(\bm{y})\ran_\tau$ is
proportional to the dipole number density $n_\tau(\x,\y)$, to be
precisely defined in Sect. 4. Here and in what follows, we use the
simpler notation $\rho_{\x}^a\equiv \rho^a_\infty({\x})$.
Specifically :
 \be\label{rhon}\hspace*{-6mm}
 \langle \rho^a(\x)\rho^a(\y)\rangle_\tau &=& - g^2C_F\big[ n_\tau(\x,\y) +
 n_\tau(\y,\x)\big]\quad{\rm for}\quad \x \ne \y\,.\ee
Since, on the other hand, the BFKL equation for a 2--point
function is well known to emerge independently of the large--$N_c$
approximation, we expect the quantity in Eq.~(\ref{rhon}) to obey
the BFKL equation for arbitrary $N_c$. Let us check that this is
indeed the case. According to Eq.~(\ref{AVEVOLO}), we have:
  \be\label{ev2rho}
 \frac{\partial}{\partial \tau} \,
 \lan \rho^a(\x)\rho^a(\y)\ran_\tau\,=\,
\frac{-1}{16\pi^3}\int\limits_{\bm{u}\bm{v}\bm{z}}
    \mathcal{M}_{\bm{u}\bm{v}\bm{z}}    \,
     \Big\lan
\big[h_{cd}(\bm{u},\bm{v},\z), \rho_{\x}^a\rho_{\y}^a \big]
\rho_{\bm
 u}^c\rho_{\bm v}^d\Big\ran_\tau, \ee
where we have relabeled the transverse coordinates internal to
$H_{\rm BREM}$ as $\bm{u},_,\bm{v}$ and $\z$. By repeated use of
the commutation relation, we shall move $h_{cd}$ (which contains
the Wilson lines $W$) to the left and then set $W=1$
 \be
[h_{cd},\, \rho_{\x}^a\rho_{\y}^a ]&=&[h_{cd},
\rho_{\x}^a]\rho_{\y}^a+\rho_{\x}^a[h_{cd}, \rho_{\y}^a]\nn &=&
[h_{cd}, \rho_{\x}^a]\rho_{\y}^a+[h_{cd}, \rho_{\y}^a]\rho_{\x}^a+
[\, \rho_{\x}^a,\, [h_{cd},\rho_{\y}^a]\, ]\nn &\longrightarrow &
 [\, \rho_{\x}^a,\, [h_{cd},\rho_{\y}^a]\, ], \ee
where, as shown in the last line, only the double commutator must
be retained when computing the color glass average in
Eq.~(\ref{ev2rho}). Indeed
 \be
 [h_{cd},\rho^a_{\x}]\big|_{W=1}=0 \; , \label{zero} \ee
as it can be easily checked by using Eq.~(\ref{defh}) together
with the following commutator, which in turn follows from
Eq.~(\ref{PB}) ($ \delta_{\bm{u}\bm{x}}\equiv
\delta^{(2)}(\bm{u}-\bm{x})$) :
 \be\label{COMMUT1}
 [\,\rho^a_{\x}, (\wt W_{\bm{u}} \wt W^\dagger_{\bm{v}})_{cd}\,]=
 g\delta_{\bm{u}\bm{x}}\big(T^a\wt W_{\bm{u}} \wt W^\dagger_{\bm{v}}\big
 )_{cd} - g\delta_{\bm{v}\bm{x}}\big(\wt W_{\bm{u}}
 \wt W^\dagger_{\bm{v}}T^a\big )_{cd}\,.\ee
To evaluate the right hand side of Eq.~(\ref{ev2rho}) we also need
  \be\label{COMMUT2}
 \big[\,\rho^a_{\x}, [\,\rho^b_{\y}, (\wt W_{\bm{u}}
 \wt W^\dagger_{\bm{v}})_{cd}
 \,]\,\big]&=&
 g^2\delta_{\bm{u}\bm{y}}\left\{\delta_{\bm{u}\bm{x}}
 \big(T^bT^a\wt W_{\bm{u}} \wt W^\dagger_{\bm{v}}\big
 )_{cd} - \delta_{\bm{v}\bm{x}}\big(T^b\wt W_{\bm{u}}
 \wt W^\dagger_{\bm{v}}T^a\big )_{cd}\right\}\nn
 &{}&-
 g^2\delta_{\bm{v}\bm{y}}\left\{\delta_{\bm{u}\bm{x}}
 \big(T^a\wt W_{\bm{u}} \wt W^\dagger_{\bm{v}}T^b\big
 )_{cd} - \delta_{\bm{v}\bm{x}}\big(\wt W_{\bm{u}}
 \wt W^\dagger_{\bm{v}}T^aT^b\big )_{cd}\right\}\, .\ee
From now on, simple algebra yields
 \be\label{ev2rho1}
 \frac{\partial}{\partial \tau} \,
 \lan \rho^a(\x)\rho^a(\y)\ran_\tau\,=\,
 \frac{g^2N_c}{8\pi^3}\int\limits_{\bm{u}\bm{v}\bm{z}}
    \mathcal{M}_{\bm{u}\bm{v}\bm{z}}    \,
 \big(-\delta_{\bm{u}\x}\delta_{\bm{v}\y}
+\delta_{\bm{u}\x}\delta_{\z\y}+\delta_{\z\x}\delta_{\bm{v}\y}\big)
\,
 \lan \rho_{\bm{u}}^c\rho_{\bm{v}}^c \ran_\tau\,, \ee
where we have neglected the terms which vanish at $\x \neq \y$ and
used the symmetry $\bm{u} \leftrightarrow \bm{v}$ of the kernel in
the Hamiltonian. After also using Eq.~(\ref{rhon}), this is
finally rewritten as
\begin{align}\label{evolnumber}
    \frac{\del n_\tau(\bm{x},\bm{y})}{\del \tau}=
    \frac{\abar}{2\pi} \int_{\bm{z}}\,
    \big[
    -&\, {\cal M} ({\bm{x}},{\bm{y}},{\bm{z}})\,n_\tau(\bm{x},\bm{y})
    \nonumber \\
    +&\, {\cal M} ({\bm{x}},{\bm{z}},{\bm{y}})\,n_\tau(\bm{x},\bm{z})
    + {\cal M} ({\bm{z}},{\bm{y}},{\bm{x}})\,n_\tau(\bm{z},\bm{y})
    \big]
\end{align}
(with $\bar{\alpha}_s=\alpha_sN_c/\pi$), which at large $N_c$ is
recognized as the BFKL equation for the dipole number density
\cite{LL04,IT04}, but which is valid as written for arbitrary
$N_c$.

Note that the same equation could have been obtained by first
expanding the Wilson lines in the Hamiltonian (\ref{HBREM}) in a
power series in derivatives, then keeping the first non--trivial
terms (the second order ones) in this expansion to deduce the BFKL
Hamiltonian:
 \be\label{HBFKL}
    H_{\rm BFKL} & = &  {{-g^2} \over {16\pi^3}}
\int\limits_{\bm{u}\bm{v}\bm{z}}{\cal M}_{\bm{u}\bm{v}\bm{z}}
f^{ace}f^{bde}  \left[ {\delta \over {\delta \rho^c({\bm{u}})}}-
{\delta \over {\delta \rho^c(\bm{z})}}\right]
 \left[
{\delta \over {\delta \rho^d(\bm{z})}} - {\delta \over {\delta
\rho^d(\bm{v})}}\right]\rho^a(\bm{u})
\rho^b(\bm{v}), \nonumber \\
 & & \ee
and finally using this Hamiltonian in the evolution equation for
$\lan \rho^a(\bm{x}) \rho^a(\bm{y})\ran_\tau$ : 
 \be\label{ev2rho2}
 \frac{\partial}{\partial \tau} \,
 \lan \rho^a(\x)\rho^a(\y)\ran_\tau\,=\,\int D[\rho]\,
 \rho^a(\x)\rho^a(\y) \big(-\,
 H_{\rm BFKL}\,Z_{\tau}[\rho]\big)\,.\ee
After some integration by parts, the functional derivatives in
Eq.~(\ref{HBFKL}) are brought to act on the factors of $\rho$ in
the operator, and then Eq.~(\ref{ev2rho1}) immediately follows.

One may think that the first derivation of Eq.~(\ref{ev2rho1}), in
which the Wilson lines inside $H_{\rm BREM}$ were kept unexpanded,
is more general than the one based on the BFKL Hamiltonian
(\ref{HBFKL}), but this is only illusory: Since ${\cal O}=
\rho^a_{\x}\rho^a_{\y}$ is quadratic in $\rho$, all the higher
order $\rho$--derivatives beyond the second--order ones kept in
Eq.~(\ref{HBFKL}) do not contribute to its evolution. Thus, in so
far as Eq.~(\ref{ev2rho1}) is concerned, the two methods presented
above --- the use of the commutation relations and the derivative
expansion of the BREM Hamiltonian --- are equivalent with each
other. However, for more general situations (e.g., more
complicated observables, or the evolution of the weight function
that we shall consider in the next section), the use of the
commutation relations turns out to be more convenient as it avoids
potential ambiguities with the $x^+$--ordering of the functional
derivatives in the expansion of the Wilson line (\ref{Wdef}).

\section{Dipole picture from the Bremsstrahlung Hamiltonian}
\setcounter{equation}{0} 

In this section we shall derive the dipole picture from the action
of the Bremsstrahlung Hamiltonian in the large--$N_c$ limit. More
precisely, we shall show that, if one starts with a single color
dipole --- a quark--antiquark pair with the quark at $\bm{u}_0$
and the antiquark at $\bm{v}_0$ --- at the initial rapidity
$\tau_0=0$, then the partonic system produced at some higher
rapidity $\tau$ through the evolution described by $H_{\rm BREM}$
at large $N_c$ can be itself characterized as a collection of
$q\bar q$ color dipoles (an {\em `onium'}), which evolves through
dipole splitting. Besides the Bremsstrahlung Hamiltonian
(\ref{HBREM}), the crucial ingredient in this picture is the
description of the onium wavefunction as a {\em color glass},
which in turn requires the proper definition of the dipole
operator as a {\em color source}.

The first formulation of the onium as a color glass has been given
in Ref. \cite{IM031}, under the assumption that each dipole is a
color source for only two gluons. The evolution of the associated
weight function through gluon splitting has been exhibited in Ref.
\cite{IM031}, but it was only later, in Ref. \cite{IT04}, that one
has realized that this evolution cannot be fully accounted for by
the JIMWLK Hamiltonian. The appropriate Hamiltonian has been
constructed shortly after, by Mueller, Shoshi and Wong
\cite{MSW05}. It is the sum of the BFKL Hamiltonian (\ref{HBFKL})
plus a term involving four $\rho$--derivatives which describes
dipole splitting (again, under the assumption that each dipole can
radiate only two gluons). In what follows, we shall generalize the
construction in Refs. \cite{IM031,MSW05} to the case where a
dipole can radiate arbitrarily many (small--$x$) gluons. That is,
we shall construct the corresponding dipole operator and onium
weight function, and show that the Bremsstrahlung Hamiltonian
(\ref{HBREM}) is the appropriate generalization of the MSW
Hamiltonian \cite{MSW05}.

We start with a brief summary of the results in Refs.
\cite{IM031,MSW05}, to which we shall refer as the {\em `Dipole
Model'} (DM). The onium (color--glass) weight function in the DM
reads: \be
 \label{ZDM}
 Z_\tau^{\rm DM}[\rho ] \; = \;
 \sum_{N=1}^{\infty} \, \int d\Gamma_N \,
 P_N(\{\bm{z}_i\};\tau)
 \, \prod_{i=1}^{N} D_0^{\dagger}(\bm{z}_{i-1},\bm{z}_i)
  \, \delta[\rho] \; , \ee
where $P_N(\{\bm{z}_i\};\tau)$ denotes the probability density to
find a given configuration of $N$ dipoles at rapidity $\tau$ (the
configuration being specified by $N-1$ transverse coordinates
$\{\bm{z}_i\}=\{\bm{z}_1, \bm{z}_2,...\bm{z}_{N-1}\}$, such that
the coordinates of the $N$ dipoles are $(\bm{z}_0,\bm{z}_1)$,
$(\bm{z}_1,\bm{z}_2)$,...,$(\bm{z}_{N-1},\bm{z}_N)$, with
$\bm{z}_0 \equiv \bm{u}_0$ and $\bm{z}_N \equiv \bm{v}_0$), and
$d\Gamma_N$ denotes the measure for the phase--space integration:
$d{\Gamma}_N\,=\,{\rm d}^2\bm{z}_1{\rm d}^2\bm{z}_2\dots{\rm
d}^2\bm{z}_{N-1}$. Furthermore, $D^{\dagger}_0(\bm u,\bm v)$ is
the DM dipole creation operator, with the dipole assimilated to
the source of two gluons:
  \be
 \label{DCDM}
 D^{\dagger}_0(\bm x,\bm y)\,\equiv\,
 1 + \frac{g^2}{4 N_c}
\Big(\frac{\delta}{\delta \rho^a(\bm x)} -\frac{\delta}{\delta
 \rho^a(\bm y)} \Big)^2\,, \ee
where $ \rho^a(\bm x) = \int dx^- \rho^a(x^-,\bm x)$ should be
interpreted as the color charge density in the transverse plane at
the interaction time. (There is no explicit $x^+$--dependence in
the DM picture; see also Eq.~(\ref{intdrho}) below.) Finally, the
delta--functional $\delta[\rho]\equiv \delta[\rho^a(\bm x)]$ is
defined in the context of the functional integral over $\rho^a(\bm
x)$.

The evolution of the DM is driven by
 \be \label{RGEDM}
 \frac{\partial}{\partial \tau} \,Z_{\tau}^{\rm DM}[\rho]\,=\,-\,
 H_{\rm MSW}\,Z_{\tau}^{\rm DM}[\rho]\,,\ee
with the Mueller--Shoshi--Wong Hamiltonian
 \be \label{HMSW}
 H_{\rm MSW}\,=\,-\,\frac{\bar\alpha_s}{2\pi} \int\limits_{\bm{x}\bm{y}\bm{z}}
{\cal M}(\bm x,\bm y,\bm z) \Big[-D^{\dagger}_0(\bm x,\bm y)
+D^{\dagger}_0(\bm x,\bm z) D^{\dagger}_0(\bm z,\bm y) \Big]
 D_0(\bm x,\bm y),\ee
where $D_0(\bm x,\bm y)$ is the dipole annihilation operator
within the DM:
 \be \label{DADM}
D_0(\bm{x},\bm{y}) \; = \; -\frac{1}{g^2 N_c} \,\rho^a(\bm x)
 \rho^a(\bm y) \quad {\rm for}\quad \bm x \ne \bm y .\ee
One can check indeed that:
 \be \label{DADCDM}
 [\,D_0(\bm{x},\bm{y}),D^{\dagger}_0(\bm u,\bm v)\,]\,
 \approx\,\frac{1}{2}
 \big(\delta_{\bm{u}\bm{x}}\delta_{\bm{v}\bm{y}} +
 \delta_{\bm{u}\bm{y}}\delta_{\bm{v}\bm{x}}\big),\ee
where the approximate equality sign means that the equality holds
in the large--$N_c$ limit. (In Sect. 4, we shall demonstrate a
relation similar to Eq.~(\ref{DADCDM}) in a more general context.)

By inserting Eqs.~(\ref{ZDM}) and (\ref{HMSW}) into
Eq.~(\ref{RGEDM}), then using Eq.~(\ref{DADCDM}) to successively
commute the annihilation operator $D_0(\bm x,\bm y)$ from $H_{\rm
MSW}$ to the right of the creation operators
$D_0^{\dagger}(\bm{z}_{i-1},\bm{z}_i)$ from $Z_\tau^{\rm DM}$, and
finally using the fact that $D_0\delta[\rho] =0$, one finds the
following evolution equation for the onium weight function:
\be\frac{\partial}{\partial \tau} \,Z_{\tau}^{\rm DM}[\rho]\, &\,
 \approx\,&
\frac{\bar{\alpha}_s}{2\pi} \sum_{N=1}^{\infty} \, \int d\Gamma_N
\, P_N(\tau)
    \sum_{i=1}^N
\int\limits_{\bm{z}}
    \mathcal{M}(\bm{z}_{i-1},\bm{z}_i,\bm{z})
    \nn && \quad\times \big[
    D^{\dagger}_0(\bm{z}_{i-1},\bm{z})
    D^{\dagger}_0(\bm{z},\bm{z}_i)
    -D^{\dagger}_0(\bm{z}_{i-1},\bm{z}_i)
    \big]  \prod_{j\neq i}
     D^{\dagger}_0(\bm{z}_{j-1},\bm{z}_j) \,
     \delta[\rho], \label{EVZDM}
 \ee
which is the expression of the dipole picture in the color glass
representation \cite{IM031}: the evolution of the weight function
proceeds through dipole splitting. From here on, one can proceed
in the standard way \cite{IM031,LL04,IT04} to deduce evolution
equations for the dipole densities and probabilities (see the
discussion at the end of this section and in Sect. 4).

We now return to the general case where a dipole can emit an
arbitrary number of gluons, and show that an evolution equation
similar to Eq.~(\ref{EVZDM}) (which is synonymous of the dipole
picture) is obtained also in that case provided we replace $H_{\rm
MSW}$ by $H_{\rm BREM}$ and the `bare dipole' creation operator
$D^{\dagger}_0$ by the following, `dressed dipole', operator
\cite{KL4} :
 \be
 D^{\dagger}(\bm{x},\bm{y}) \,= \,\frac{1}{N_c} \,\tr \big(
 W(\bm{x})W^{\dagger}({\bm y})\big), \label{ii} \ee
where the Wilson lines are in the fundamental representation ($W$
stays for the quark, and $W^{\dagger}$ for the antiquark). To
second order in the expansion of the Wilson lines in powers of
$\delta/\delta\rho$, and with the following identification
  \be\label{intdrho}
\frac{\delta}{\delta \rho^a(\bm x)}\,\equiv\,\int dx^+\,
 \frac{\delta}{\delta \rho^a(x^+,\bm x)}\,,
 \ee
Eq.~(\ref{ii}) reduces to the two--gluon emission operator,
Eq.~(\ref{DCDM}). In general, Eq.~(\ref{ii}) can be seen as the
gauge--invariant generalization of Eq.~(\ref{DCDM}) to the regime
where the derivatives are formally strong : $g\int dx^+
(\delta/\delta\rho)\sim 1$ (cf. the discussion after
Eq.~(\ref{Wdef})).

The onium weight function will be now constructed by analogy with
Eq.~(\ref{ZDM}). That is, we start by assuming that the weight
function $Z_\tau[\rho]$ in the large--$N_c$ limit can be cast in
the following, dipolar, form :
 \be\label{ZDipole}
 Z_\tau[\rho ] \; = \;
 \sum_{N=1}^{\infty} \, \int d\Gamma_N \,
 P_N(\{\bm{z}_i\};\tau)
 \, \prod_{i=1}^{N} D^{\dagger}(\bm{z}_{i-1},\bm{z}_i)
  \, \delta[\rho] \; , \ee
and then show that this particular structure is indeed preserved
by the evolution under the action of the Bremsstrahlung
Hamiltonian (\ref{HBREM}) and for large $N_c$.

The subsequent mathematical manipulations can perhaps be better
understood if one notices their analogy (in fact, {\em duality})
to manipulations which are by now familiar in the context of the
JIMWLK evolution (see, e.g., Refs. \cite{RGE,ODDERON}).
Specifically, the Bremsstrahlung and JIMWLK evolutions are known
to be {\em dual} to each other \cite{KL3,BREM}, in the sense that
the corresponding Hamiltonians and also the respective Poisson
brackets get interchanged with each other under the following
duality transformations\footnote{Note that the ``$\infty$"
subscript in $\alpha_\infty$ refers to $x^-$, and not to $x^+$;
see Ref. \cite{BREM} for details.}  :
 \be\label{DUAL}
 \frac{1}{i} \frac{\delta}{\delta
\alpha_\infty^a({\x})}\,\longleftrightarrow \,\rho^a_\infty(\x)
 \,,\qquad
 V^\dagger(\x)\longleftrightarrow \,W(\x)
 \,.\ee
Similarly,  the dipole creation operator (\ref{ii}) is dual to the
$S$--matrix operator $S(\x,\y)=\frac{1}{N_c} \,\tr (V^\dagger(\x)
V({\bm y}))$ which describes the scattering of an external dipole
in the high--density regime specific to the JIMWLK evolution (cf.
Eq.~(\ref{Tdipole})). Thus, clearly, $H_{\rm BREM}$ acts on the
{\em internal} dipoles in the same way  as $H_{\rm JIMWLK}$ does
on the {\em external} ones. We know already that the action of
$H_{\rm JIMWLK}$ on scattering operators generates the Balitsky
equations \cite{B}. At large $N_c$ and for projectiles built with
dipoles, the Balitsky equations close in the space of dipole
operators, and are consistent with the dipole picture for the {\em
projectile} wavefunction \cite{JP04,LL04,IT04}. This lets us
anticipate that the action of $H_{\rm BREM}$ on the dipolar weight
function (\ref{ZDipole}) at large $N_c$ should similarly generate
the dipole picture for the {\em target} wavefunction. This will be
verified explicitly in what follows.

We thus need to evaluate
\begin{eqnarray} \frac{\partial}{\partial \tau} \,Z_{\tau}[\rho]\,
&=& -\,H_{\rm BREM}\,Z_{\tau}[\rho]\,
 \nonumber \\ &=& \sum_{N=1}^{\infty} \, \int d\Gamma_N \,
P_N(\tau)\frac{1}{16\pi^3}\,
    \int\limits_{\bm{x}\bm{y}\bm{z}}
    \mathcal{M}_{\bm{x}\bm{y}\bm{z}}\,
    h_{ab}
   \rho^a_{\bm{x}} \rho^b_{\bm{y}}\prod_{i=1}^{N}
   D^{\dagger}(\bm{z}_{i-1},\bm{z}_i) \,
   \delta[\rho]
 \end{eqnarray}
in the large--$N_c$ limit. For more clarity, it is convenient to
use the notations $\bm{u}_i\equiv \bm{z}_{i-1}$ and
$\bm{v}_i\equiv \bm{z}_i$ for the transverse coordinates of the
$i$--th dipole; it is then understood that
$\bm{u}_{i+1}=\bm{v}_i$. Let us focus on the action of $H_{\rm
BREM}$ on the weight function $Z_N$ for a given configuration of
$N$ dipoles. This involves:
\begin{eqnarray}\label{ZN}
 h_{ab}\rho^a_{\bm{x}} \rho^b_{\bm{y}} \,Z_N(\{\bm{u}_i,\bm{v}_i\})
 \,\equiv\,
    h_{ab}\rho^a_{\bm{x}} \rho^b_{\bm{y}}\,
   \prod_{i=1}^{N} D^{\dagger}(\bm{u}_i,\bm{v}_i) \,
   \delta[\rho]\,.
 \end{eqnarray}
To proceed, we need to commute $\rho_{\x}^a\rho_{\y}^b$ to the
right of the dipole creation operators and then use $\rho \,
\delta[\rho] \equiv 0$. The relevant commutator is
\begin{align}
 \big[\rho^a_{\bm{x}}\rho^b_{\bm{y}} ,
D^{\dagger}_{\bm{u}\bm{v}}\big] \; = \;
 \big[\rho^a_{\bm{x}}, D^{\dagger}_{\bm{u}\bm{v}} \big] \,
\rho^b_{\bm{y}} +
 \big[\rho^b_{\bm{y}}, D^{\dagger}_{\bm{u}\bm{v}} \big] \,
 \rho^a_{\bm{x}}
 +  \big[ \rho^a_{\bm{x}},\big[ \rho^b_{\bm{y}} ,
D^{\dagger}_{\bm{u}\bm{v}}\big]\big]
\end{align}
By adapting Eqs.~(\ref{COMMUT1})--(\ref{COMMUT2}) to the
fundamental representation, we obtain
 \begin{align}
 \label{eq:rrdcom}
 \big[\rho^a_{\bm{x}} \rho^b_{\bm{y}} ,
D^{\dagger}_{\bm{u}\bm{v}}\big] \; = \;
 \frac{g}{N_c} (\delta_{\bm{xu}}-\delta_{\bm{xv}}) \tr(t^a
W_{\bm{u}} W_{\bm{v}}^{\dagger} )
 \, \rho^b_{\bm{y}} \nonumber \\
  +\frac{g}{N_c} (\delta_{\bm{yu}}-\delta_{\bm{yv}})
  \tr(t^b W_{\bm{u}} W_{\bm{v}}^{\dagger} )
   \, \rho^a_{\bm{x}}  \nonumber \\
  + g^2 (\delta_{\bm{yu}}-\delta_{\bm{yv}}) \, \Big[
\delta_{\bm{xu}} \frac{1}{N_c}
  \tr (W_{\bm{u}} W_{\bm{v}}^{\dagger} t^b t^a) - \delta_{\bm{xv}}
\frac{1}{N_c}
  \tr (W_{\bm{u}} W_{\bm{v}}^{\dagger} t^a t^b)    \Big],
 \end{align}
where the expression in the last line represents the double
commutator term. We shall shortly verify that the first two terms
on the right hand side (proportional to $\rho$), when multiplied
by $h_{ab}$ and acting on $Z_N$, give subleading contributions at
large $N_c$ compared to the last term, which does not involve
$\rho$. Introducing the more compact notation
\begin{align}\label{ndipcom}
\big[ \rho^a_{\bm{x}} \rho^b_{\bm{y}} ,
D^{\dagger}_{\bm{u}\bm{v}}\big]_{\rm non-dipole} \; \equiv \;
 \frac{g}{N_c} (\delta_{\bm{xu}}-\delta_{\bm{xv}}) \tr(t^a
W_{\bm{u}} W_{\bm{v}}^{\dagger} )
 \, \rho^b_{\bm{y}} \nonumber \\
  +\frac{g}{N_c} (\delta_{\bm{yu}}-\delta_{\bm{yv}}) \tr(t^b
W_{\bm{u}} W_{\bm{v}}^{\dagger} )
   \, \rho^a_{\bm{x}}  \nonumber \\
   \; \equiv \;
A^a_{\bm{u}\bm{v}\bm{x}} \, \rho^b_{\bm{y}} \, + \,
B^b_{\bm{u}\bm{v}\bm{y}} \, \rho^a_{\bm{x}},
\end{align}
where $A$ and $B$ do not contain $\rho$,
we can write the result of the first commutation as follows (in
simplified notations whose meaning should be obvious) :
 \be\label{firstcom}
 \rho^a_{\bm{x}} \rho^b_{\bm{y}}
 D^{\dagger}_{1}D^{\dagger}_{2}\dots D^{\dagger}_{N}&\,=\,&
 D^{\dagger}_{1}\,\rho^a_{\bm{x}} \rho^b_{\bm{y}}\,
 D^{\dagger}_{2}\dots D^{\dagger}_{N}\,
 + \,\big[ \rho^a_{\bm{x}} \rho^b_{\bm{y}} ,
 D^{\dagger}_{\bm{u}_1\bm{v}_1}\big]_{\rm non-dipole}\,
 D^{\dagger}_{2}\dots D^{\dagger}_{N} \nn
 &{}&
 + \,\big[ \rho^a_{\bm{x}},\big[ \rho^b_{\bm{y}} ,
D^{\dagger}_{\bm{u}_1\bm{v}_1}\big]\big]\,
 D^{\dagger}_{2}\dots D^{\dagger}_{N} \,.\ee
The last, double--commutator, term in the r.h.s. of the above
equation is already independent of $\rho$, so we can operate with
$h_{ab}$ on it :
 \begin{align}\label{Bal1}
 & \frac{1}{16\pi^3}\int\limits_{\bm{x}\bm{y}\bm{z}}
    \mathcal{M}_{\bm{x}\bm{y}\bm{z}}\,
    h_{ab}(\x,\y,\z)\,\big[ \rho^a_{\bm{x}},\big[ \rho^b_{\bm{y}} ,
 D^{\dagger}_{\bm{u}_1\bm{v}_1}\big]\big]\,=\,\nn
 & \quad\qquad \,=\, -\frac{g^2}{8\pi^3 N_c}\int\limits_{\bm{z}}
 \mathcal{M}_{\bm{u}_1\bm{v}_1\bm{z}}\,
 h_{ab}(\bm{u}_1,\bm{v}_1,\bm{z})\,
 \tr (W_{\bm{u}_1} W_{\bm{v}_1}^{\dagger} t^b t^a) \,=\,\nn
 & \quad\qquad \,=\,
  \frac{\bar\alpha_s}{2\pi}\int\limits_{\bm{z}}
 \mathcal{M}_{\bm{u}_1\bm{v}_1\bm{z}}\, \Big\{-\frac{1}{N_c}
 \tr (W_{\bm{u}_1} W_{\bm{v}_1}^{\dagger}) + \frac{1}{N_c}
 \tr (W_{\bm{u}_1} W_{\bm{z}}^{\dagger})
 \,\frac{1}{N_c}
 \tr (W_{\bm{z}} W_{\bm{v}_1}^{\dagger})\Big\}
 \end{align}
where we have also used the identity
$(t^a)_{ij}(t^a)_{kl}=\frac{1}{2}\delta_{il}\delta_{jk}-\frac{1}{2N_c}
\delta_{ij}\delta_{kl}$. The last equation is dual to the first
Balitsky equation \cite{B}, and in fact this has been obtained
here through manipulations similar to those usually performed in
the derivation of the Balitsky hierarchy from the JIMWLK equation
\cite{RGE}. The first, negative, term within the braces in
Eq.~(\ref{Bal1}) describes the probability that the original
dipole $({\bm{u}_1,\bm{v}_1})$ survive without splitting, while
the second, positive, term describes the splitting of the original
dipole into the new dipoles $({\bm{u}_1,\bm{z}})$ and
$({\bm{z},\bm{v}_1})$ .

We shall now check that the non--dipolar  contribution to
Eq.~(\ref{firstcom}) is indeed suppressed at large $N_c$. To that
aim, we take one particular piece in the non--dipolar commutator
(\ref{ndipcom}), say the first piece $A^a_{\bm{u}_1\bm{v}_1\bm{x}}
\rho^b_{\bm{y}}$, and consider its action on the dipole creation
operators which appear on its right in Eq.~(\ref{firstcom}):
 \be\label{Arho1}
A^a_{\bm{u}_1\bm{v}_1\bm{x}} \, \rho^b_{\bm{y}} D_{\bm{u}_2
\bm{v}_2}^{\dagger} D_{\bm{u}_3 \bm{v}_3}^{\dagger} \dots
D_{\bm{u}_{N} \bm{v}_N}^{\dagger} \, \delta[\rho] &=&
 D_{\bm{u}_2 \bm{v}_2}^{\dagger}  A^a_{\bm{u}_1\bm{v}_1\bm{x}} \,
\rho^b_{\bm{y}}
  D_{\bm{u}_3 \bm{v}_3}^{\dagger}
\dots D_{\bm{u}_{N}\bm{v}_N }^{\dagger} \, \delta[\rho] \nonumber
\\ &+& \big[A^a_{\bm{u}_1\bm{v}_1\bm{x}} \, \rho^b_{\bm{y}} ,
D_{\bm{u}_2 \bm{v}_2}^{\dagger} \big] D_{\bm{u}_3
\bm{v}_3}^{\dagger} \dots D_{\bm{u}_{N} \bm{v}_N}^{\dagger} \,
\delta[\rho] \;  . \ee The above commutator is independent of
$\rho$, and the same is true for all the other commutators
generated when $\rho^b$ is further commuted towards the right. It
is therefore sufficient to evaluate the action of $h_{ab}$ on one
such a commutator. We have:
\begin{align}
\big[A^a_{\bm{u}_1\bm{v}_1 \bm{x}} \, \rho^b_{\bm{y}} ,
D_{\bm{u}_2 \bm{v}_2}^{\dagger} \big] = \frac{g}{N_c}
(\delta_{\bm{xu}_1}-\delta_{\bm{xv}_1}) \Tr(t^a W_{\bm{u}_1}
W_{\bm{v}_1}^{\dagger} ) \big[\rho^b_{\bm{y}} ,
D_{\bm{u}_2\bm{v}_2}^{\dagger} \big] \;
\nonumber \\
= \frac{g^2}{N_c^2} (\delta_{\bm{xu}_1}-\delta_{\bm{xv}_1})
(\delta_{\bm{yu}_2}-\delta_{\bm{yv}_2})
 \Tr(t^a W_{\bm{u}_1} W_{\bm{v}_1}^{\dagger} )  \Tr(t^b W_{\bm{u}_2}
W_{\bm{v}_2}^{\dagger}
 ). \label{above}
\end{align} Let us consider the action of the first two terms in
$h_{ab}(\bm{x},\bm{y},\bm{z})$ on this commutator:
\begin{align}
   \frac{g^2}{N_c^2} \int\limits_{\bm{x}\bm{y}\bm{z}}
    \mathcal{M}_{\bm{x}\bm{y}\bm{z}}(\delta_{\bm{xu}_1}-\delta_{\bm{xv}_1})
(\delta_{\bm{yu}_2}-\delta_{\bm{yv}_2})\big(1+\widetilde{W}_{\bm
x}\widetilde{W}_{\bm y}^{\dagger}\big)_{ab} \Tr(t^a W_{\bm{u}_1}
W_{\bm{v}_1}^{\dagger} ) \Tr(t^b
W_{\bm{u}_2}W_{\bm{v}_2}^{\dagger}) \nn =\frac{g^2}{N_c}
\int\limits_{\bm{z}}\big(\mathcal{M}_{\bm{u}_1\bm{u}_2\bm{z}}-
\mathcal{M}_{\bm{u}_1\bm{v}_2\bm{z}}
-\mathcal{M}_{\bm{v}_1\bm{u}_2\bm{z}}+
\mathcal{M}_{\bm{v}_1\bm{v}_2\bm{z}}\big)
\Big[\frac{1}{2N_c}\tr(W_{\bm{u}_1}W_{\bm{v}_1}^{\dagger}W_{\bm{u}_1}W_{\bm{v}_2}^{\dagger})
\nonumber \\ +
\frac{1}{2N_c}\tr(W_{\bm{v}_1}^{\dagger}W_{\bm{u}_1}W_{\bm{v}_2}^{\dagger}W_{\bm{u}_2})
-\frac{1}{N_c}\tr(W_{\bm{u}_1}W_{\bm{v}_1}^{\dagger})
 \,\frac{1}{N_c}\tr(W_{\bm{u}_2}W_{\bm{v}_2}^{\dagger})\Big].
\label{ww}
\end{align}
The r.h.s. of Eq.~(\ref{ww}) is of order $\bar{\alpha}_s/N_c^2$,
and hence it is suppressed by a factor of $1/N_c^2$ compared to
the dipole contribution in Eq.~(\ref{Bal1}). Similarly, one can
verify that the other contributions, due to the last two terms in
$h_{ab}(\bm{x},\bm{y},\bm{z})$, are also suppressed. Once again,
this property has a dual counterpart in the context of the
Balitsky--JIMWLK equations: The non--dipolar terms in the
evolution equation for the scattering amplitude of a projectile
made with {\em two} dipoles are suppressed at large $N_c$.

Retaining only the dipolar contribution in Eq.~(\ref{Bal1}), we
finally obtain the following evolution equation for the weight
function in the large--$N_c$ limit:
\begin{eqnarray} \frac{\partial}{\partial \tau} \,Z_{\tau}[\rho]\,
&=& \frac{\bar{\alpha}_s}{2\pi} \sum_{N=1}^{\infty} \, \int
d\Gamma_N \, P_N(\tau)
    \sum_{i=1}^N
\int\limits_{\bm{z}}
    \mathcal{M}({\bm{u}_i,\bm{v}_i,\bm{z}})\nn && \qquad\times \big[
    D^{\dagger}(\bm{u}_{i},\bm{z})\,D^{\dagger}(\bm{z},\bm{v}_{i})
    -D^{\dagger}(\bm{u}_{i},\bm{v}_{i})\big]  \prod_{j\neq i}
     D^{\dagger}(\bm{u}_{j},\bm{v}_{j}) \,
     \delta[\rho]. \label{eq:BalNc2}
\end{eqnarray}
As anticipated, this has the same structure as the corresponding
equation in the Dipole Model, Eq.~(\ref{EVZDM}), except for the
replacement of the creation operator for a {\em bare} dipole
$D^{\dagger}_0$ with the corresponding operator for a {\em
dressed} dipole $D^{\dagger}$. Note that the non--commutativity of
the color charge operators $\rho^a$ (cf. the second equation
(\ref{PB})) did not play any role in the manipulations leading to
Eq.~(\ref{eq:BalNc2}). This suggests that the ordering of the
operators should be irrelevant in the large--$N_c$ limit (at
least, for the dipole--related variables). This conclusion will be
further supported by the developments in Sect. 4.

From Eq.~(\ref{eq:BalNc2}), the dipole picture of the target
wavefunction can be developed along the same lines as in Ref.
\cite{IM031}. Namely, Eq.~(\ref{eq:BalNc2}) is consistent with the
dipolar structure of the weight function, Eq.~(\ref{ZDipole}),
provided the probability densities $P_N$ which enter the latter
obey the following {\em Master equation} :
 \begin{align}\label{evolP}
    \frac{\del P_N(\bm{z}_1,...\bm{z}_{N-1};\tau)}{\del \tau}
    =&-\frac{\abar}{2\pi}
    \left[\,
    \sum_{i=1}^N \int \!d^2 \bm{z}\,
    \mathcal{M}(\bm{z}_{i-1},\bm{z}_i,\bm{z})
    \,\right]
    P_N(\bm{z}_1,...\bm{z}_{N-1};\tau)
    \nonumber \\
    &+\frac{\abar}{2\pi}
    \sum_{i=1}^{N-1}
    \mathcal{M}(\bm{z}_{i-1},\bm{z}_{i+1},\bm{z}_i)
    \,P_{N-1}
    (\bm{z}_1,...,\bm{z}_{i-1},\bm{z}_{i+1},...,\bm{z}_{N-1};\tau).
\end{align}
The first term in the r.h.s., proportional to $P_N$, is the {\em
loss term}, which describes the splitting of one dipole from the
original configuration of $N$ dipoles. The other terms,
proportional to $P_{N-1}$, are {\em gain terms} showing the
formation of the $N$--dipole configuration of interest through the
splitting of one dipole in original configurations of $N-1$
dipoles.

One can summarize the previous discussion as follows: At large
$N_c$, the Bremsstrahlung Hamiltonian (\ref{HBREM}) acts on the
onium weight function built with {\em dressed} dipoles,
Eq.~(\ref{ZDipole}), in the same way as the Dipole Model
Hamiltonian $H_{\rm MSW}$ acts on the weight function (\ref{ZDM})
built with {\em bare} dipoles. But this does not imply that
$H_{\rm MSW}$ can be deduced from $H_{\rm BREM}$ by somehow taking
the large--$N_c$ limit of the latter. That is, the derivative
expansion of $H_{\rm BREM}$, Eq.~(\ref{HMBREM}), up to fourth
order in $\delta/\delta\rho$ is neither equivalent to the
expression (\ref{HMSW}) of $H_{\rm MSW}$, nor it reduces to it in
some suitable large--$N_c$ limit. In fact, since the expressions
for the onium weight function in the Dipole Model,
Eq.~(\ref{ZDM}), and in the general Bremsstrahlung case,
Eq.~(\ref{ZDipole}), differ from each other by terms of order
$(\delta/\delta\rho)^3$ or higher, the respective Hamiltonians
{\em must} differ from each other too (already at order
$(\delta/\delta\rho)^3$), in order to yield identical results when
acting on different dipole operators. This observation explains,
in particular, why the analysis in Ref. \cite{KL4}, which has
limited itself to the extraction of the four derivative terms in
the Bremsstrahlung Hamiltonian (\ref{HBREM}), has met with
difficulties when trying to generate the dipole picture from
$H_{\rm BREM}$.

The above discussion finds its dual counterpart in the analysis of
the JIMWLK evolution in Ref. \cite{BIIT05} : At large $N_c$,
$H_{\rm JIMWLK}$ acts on the dipole operator
$S(\x,\y)=\frac{1}{N_c} \,\tr (V^\dagger(\x) V({\bm y}))$, which
describes multiple scattering, in the same way as the `Pomeron'
Hamiltonian obtained \cite{BIIT05} as the dual partner of the MSW
Hamiltonian (\ref{HMSW}) acts on the `bare' operator $S_0(\x,\y) =
1 - \frac{g^2}{4N_c}\, (\alpha_a({\x})-\alpha_a({\y}))^2$, which
describes single scattering via two gluon exchange (cf.
Eq.~(\ref{Tdipole})). The `Pomeron' Hamiltonian at high density
and, respectively, the Dipole Model Hamiltonian (\ref{HMSW}) in
the dilute regime are just {\em effective} Hamiltonians, which
generate the correct evolution equations  at large $N_c$ (for
dipole scattering amplitudes in the first case, and for target
dipole densities in the latter), but by acting in a simplified
Hilbert space in which each dipole is characterized by the
exchange of only two gluons.

It is finally interesting to notice the different ways how the
actual QCD Hamiltonians (BREM and JIMWLK) and their effective
counterparts (MSW and, respectively, Pomeron) act in the
corresponding Hilbert spaces. By construction, $H_{\rm MSW}$
implements the dipole evolution in the most straightforward way:
when acting on a collection of (bare) dipoles, it first
annihilates one dipole, which is then replaced either by the same
dipole, or by a pair of dipoles with one common leg. One may
naively expect the general Hamiltonian for dipole evolution to
have the same structure as $H_{\rm MSW}$, but in terms of {\em
dressed} (creation and annihilation) operators. However, even at
large $N_c$, the actual dipole evolution in QCD proceeds in a more
subtle way: The bilocal operator $\rho^a_{\bm{x}} \rho^b_{\bm{y}}$
within $H_{\rm BREM}$ does not annihilate a dipole, rather it
rotates this twice, around different directions in color space
(see the last line in Eq.~(\ref{eq:rrdcom})). These rotations are
then compensated by the remaining operators in the Hamiltonian
(those involving Wilson lines), which can either restore back the
original dipole, or split it into two new dipoles (cf.
Eq.~(\ref{Bal1})). Thus, although the initial and final states
--- prior to, and after the evolution --- are colorless dipoles,
the evolution proceeds through {\em colorful} configurations at
intermediate steps.

\section{Evolution equation for the dipole densities}

In the previous section, we have expressed the evolution of the
dipole picture in two equivalent ways, which both deal with the
ensemble of dipole correlations in the onium: \texttt{(i)} as a
renormalization group equation for the color glass weight function
$Z_{\tau}[\rho]$, Eq.~(\ref{eq:BalNc2}), and \texttt{(ii)} as a
Master equation for the dipole probability densities
$P_N(\{\bm{z}_i\};\tau)$, Eq.~(\ref{evolP}). Alternatively, the
same physical evolution can be expressed as an hierarchy of
ordinary evolution equations for the dipole $k$--body densities
$n^{(k)}_\tau$ ($k\ge 1$), which in principle can be derived from
any of the `functional' evolutions alluded to above provided one
knows the corresponding expressions for the dipole densities. The
derivation based on the Master equation (\ref{evolP}) has been
presented in Refs. \cite{LL04,IT04}. In what follows, we shall
derive the same equations from the evolution (\ref{eq:BalNc2}) of
the color glass. This requires to properly identify the operator
expressing the dipole number density in the
$\rho$--representation.

For more clarity, let us first recall the {\em abstract}
definition of the dipole number operator, where by `abstract' we
mean an operator which is independent of $\rho$, and thus of the
way in which we measure the dipole distribution. The abstract
dipole number density operator reads (for a $N$--dipole
configuration) \cite{IM031,IT04}
\begin{equation}\label{densityN}
    n_{N}(\bm{x},\bm{y}; \{\bm{z}_i\})=
    \sum_{i=1}^N
    \delta^{(2)}(\bm{z}_{i-1}-\bm{x})
    \delta^{(2)}(\bm{z}_i-\bm{y}),
\end{equation}
and its expectation value $n_\tau(\bm{x},\bm{y})\equiv \lan
n(\bm{x},\bm{y})\ran_\tau$ is obtained as
\begin{align}\label{avn}
 n_\tau(\bm{x},\bm{y})=\sum_{N=1}^\infty
\int d\Gamma_N \,P_N(\{\bm{z}_i\};\tau) \, n_N(\bm{x},\bm{y};
\{\bm{z}_i\}).
\end{align}
By taking a derivative with respect to $\tau$ in the above
equation and using the Master equation (\ref{evolP}), one can
check that $n_\tau(\bm{x},\bm{y})$ obeys the BFKL equation
(\ref{evolnumber}). Furthermore, the abstract operator expressing
the dipole {\em pair} density is defined as \cite{IT04}
\begin{align}\label{n2def1}
    n_N^{(2)}(\bm{x}_1,\bm{y}_1 ; \bm{x}_2,\bm{y}_2)=\!
    \sum_{\substack{j,k=1\\j \neq k}}^N
    \delta^{(2)}(\bm{z}_{j-1}-\bm{x}_1)
    \delta^{(2)}(\bm{z}_j-\bm{y}_1)
    \delta^{(2)}(\bm{z}_{k-1}-\bm{x}_2)
    \delta^{(2)}(\bm{z}_k-\bm{y}_2),
\end{align}
where the sum is restricted to different pairs of dipoles since we
do not want to count the same dipole twice. The equation obeyed by
the corresponding expectation value
$n^{(2)}_\tau(\bm{x}_1,\bm{y}_1;\bm{x}_2,\bm{y}_2)$ has been
obtained from the Master equation (\ref{evolP}) in Refs.
\cite{LL04,IT04} (see Eq.~(5.15) in Ref.~\cite{IT04}), and will be
rederived below from Eq.~(\ref{eq:BalNc2}). Abstract $k$--body
dipole density operators $n^{(k)}_N$ with $k\ge 1$ can be
similarly defined \cite{LL04,IT04}.

Let us now turn to the representation of these dipole operators in
the color glass formalism, where a dipole is described as a source
for small--$x$ gluons. Within the Dipole Model, where a dipole can
radiate only two gluons, the dipole number densities are built
with the `bare dipole' annihilation operator introduced in
Eq.~(\ref{DADM}). Specifically, by using the commutation relation
(\ref{DADCDM}), one finds (for large $N_c$ and $\x \ne \y$) :
 \be\hspace*{-0.3cm}
 D_0(\bm{x},\bm{y})\,Z_{\tau}^{\rm DM}[\rho]\, &\,
 \approx\,&
 \sum_{N=1}^{\infty} \, \int d\Gamma_N \, P_N(\tau)
    \sum_{i=1}^N
\frac{1}{2}
 \big(\delta_{\bm{z}_{i-1}\bm{x}}\delta_{\bm{z}_{i}\bm{y}} +
 \delta_{\bm{z}_{i-1}\bm{y}}\delta_{\bm{z}_{i}\bm{x}}\big)
 \prod_{j\neq i}
     D^{\dagger}_0(\bm{z}_{j-1},\bm{z}_j) \,
     \delta[\rho].\nn \label{D0DM}
 \ee
After averaging over $\rho$, which is tantamount to replacing
$D^{\dagger}_0\to 1$ (since the functional derivatives within
$D^{\dagger}_0$ give zero after integration by parts), this
yields:
 \be\label{AVD0}
 \lan D_0(\bm{x},\bm{y})\ran_\tau &\equiv & \int D[\rho]\,
 D_0(\bm{x},\bm{y})\,Z_{\tau}^{\rm DM}[\rho]\,\nn &\approx &\,
 \sum_{N=1}^{\infty} \, \int d\Gamma_N \, P_N(\tau)
    \sum_{i=1}^N
\frac{1}{2}
 \big(\delta_{\bm{z}_{i-1}\bm{x}}\delta_{\bm{z}_{i}\bm{y}} +
 \delta_{\bm{z}_{i-1}\bm{y}}\delta_{\bm{z}_{i}\bm{x}}\big)\nn
 &\equiv & (1/2)\big(n_\tau(\bm{x},\bm{y}) + n_\tau(\bm{y},\bm{x})
 \big),\ee
where in writing the last line we have recognized the average
dipole number density according to Eq.~(\ref{densityN}). Note that
the measure of the dipole density provided by the charge operator
$D_0$ is symmetrized between the quark and antiquark legs of the
dipole. One can similarly check that, at large $N_c$,
 \be\label{AVD02}
 \lan D_0(\bm{x_1},\bm{y_1})\,D_0(\bm{x_2},\bm{y_2})\ran_\tau &\,=\,&
 \frac{1}{4}\,\big(n^{(2)}_\tau(\bm{x}_1,\bm{y}_1;\bm{x}_2,\bm{y}_2)+
n^{(2)}_\tau(\bm{y}_1,\bm{x}_1;\bm{x}_2,\bm{y}_2)\nonumber
\\ &{}&\quad +\,n^{(2)}_\tau(\bm{x}_1,\bm{y}_1;\bm{y}_2,\bm{x}_2)
 +n^{(2)}_\tau(\bm{y}_1,\bm{x}_1;\bm{y}_2,\bm{x}_2)\big),\ee
so long as we restrict ourselves to {\em different} dipoles; that
is, we exclude configurations such that the two measured dipoles
are identical with each other: $\{\bm{x_1}=\bm{x_2};\,
\bm{y_1}=\bm{y_2}\}$ or $\{\bm{x_1}=\bm{y_2};
\,\bm{y_1}=\bm{x_2}\}$. Besides, we exclude, as usual, the
zero--size dipoles; that is, we assume $\bm{x_1}\ne\bm{y_1}$ and
$\bm{x_2}\ne\bm{y_2}$.

We thus see that, within the DM, the (bare) dipole {\em
annihilation} operator plays also the role of a {\em number}
operator. This degeneracy is possible because our present use of
creation and annihilation operators is somewhat different from
their standard use in quantum mechanics: Whereas the weight
function (\ref{ZDM}) plays naturally the role of a ``ground
state'' for the (bare) dipole ``Fock space", the color glass
expectation value on this `ground state' involves $Z_{\tau}^{\rm
DM}$ itself, and not $|Z_{\tau}^{\rm DM}|^2$.

The above identification of $D_0$ with the dipole number operator
is also consistent with the color glass evolution of the DM, as
encoded in Eq.~(\ref{EVZDM}) for $Z_{\tau}^{\rm DM}$. Using the
latter or, more directly, acting with the MSW Hamiltonian
(\ref{HMSW}) on the relevant operators built with $D_0$, one can
indeed check that the correlations of $D_0$ introduced in
Eqs.~(\ref{AVD0}) and (\ref{AVD02}) obey the evolution equations
expected for the respective dipole densities \cite{LL04,IT04}. The
most efficient way to perform this calculation is to use the
representation of $H_{\rm MSW}$ as a fourth--order differential
operator (cf. Eq.~(\ref{DCDM})) acting on functionals of
$\rho^a(\x)$.

Let us now turn to the most interesting case, in which a dipole is
allowed to radiate arbitrarily many gluons and the respective
creation operator involves Wilson lines, cf. Eq.~(\ref{ii}). In
the previous manipulations leading to the dipole picture, we did
not need to introduce the corresponding annihilation operator,
because the natural candidate in that respect --- the bilocal
operator $\rho^a(\bm x) \rho^a(\bm y)$ --- does not enter the
Bremsstrahlung Hamiltonian (unlike what happens in the Dipole
Model). Yet, as we shall now explain, this operator plays an
important role also in the  general case, not as an annihilation
operator, but rather as a {\em dipole number operator}.
Specifically, if one defines
 \be \label{DA}
 D(\bm{x},\bm{y}) \; \equiv \; -\frac{1}{g^2 N_c} \,\rho^a_{\infty}(\bm x)
 \rho^a_{\infty}(\bm y) \quad {\rm for}\quad \bm x \ne \bm y ,\ee
which differs from the corresponding DM operator,
Eq.~(\ref{DADM}), only by the presence of the time argument
$x^+=\infty$ (that we have temporarily reintroduced for more
clarity), then this operator is gauge invariant and obeys the
following commutation relation at large $N_c$
 \be \label{DADCDNC}
 [\,D(\bm{x},\bm{y}),D^{\dagger}(\bm u,\bm v)\,]\,
 \approx\,\frac{1}{2}
 \big(\delta_{\bm{u}\bm{x}}\delta_{\bm{v}\bm{y}} +
 \delta_{\bm{u}\bm{y}}\delta_{\bm{v}\bm{x}}\big)\,
 D^{\dagger}(\bm u,\bm v),\ee
which qualifies it as a number operator, as anticipated. To verify
the above commutator, note that, for $\x\ne\y$,
Eq.~(\ref{eq:rrdcom}) yields
\begin{align}
 \label{DDCOM}
 \big[\, D_{\bm{x}\bm{y}} ,
D^{\dagger}_{\bm{u}\bm{v}}\,\big] \; = \;
 -\frac{1}{gN_c^2} (\delta_{\bm{xu}}-\delta_{\bm{xv}}) \tr(t^a
W_{\bm{u}} W_{\bm{v}}^{\dagger} )
 \, \rho^a_{\bm{y}} \nonumber \\
  -\frac{1}{gN_c^2}(\delta_{\bm{yu}}-\delta_{\bm{yv}})
  \tr(t^a W_{\bm{u}} W_{\bm{v}}^{\dagger} )
   \, \rho^a_{\bm{x}}  \nonumber \\
  +\, \frac{1}{2}\,\big(\delta_{\bm{u}\bm{x}}\delta_{\bm{v}\bm{y}} +
 \delta_{\bm{u}\bm{y}}\delta_{\bm{v}\bm{x}}\big)\,
 D^{\dagger}_{\bm{u}\bm{v}}, \end{align}
where only the last term in the r.h.s. survives when acting on the
onium weight function (\ref{ZDipole}) and at large $N_c$. (This
can be checked through manipulations similar to those in
Eqs.~(\ref{Arho1})--(\ref{ww}).)

Eq.~(\ref{DADCDNC}) implies that $Z_N$ --- the weight function for
a given configuration of $N$ dipoles; see Eq.~(\ref{ZN}) --- is an
eigenstate of $D$ at large $N_c$ (compare to Eq.~(\ref{D0DM})) :
\be\label{DZN}
 D(\bm{x},\bm{y})\,\,Z_N(\{\bm{u}_i,\bm{v}_i\})\,\approx\,
 \frac{1}{2}\sum_{i=1}^N
 \big(\delta_{\bm{z}_{i-1}\bm{x}}\delta_{\bm{z}_{i}\bm{y}} +
 \delta_{\bm{z}_{i-1}\bm{y}}\delta_{\bm{z}_{i}\bm{x}}\big)
 \,Z_N(\{\bm{u}_i,\bm{v}_i\}),\ee
which in turn implies (cf. Eq.~(\ref{avn})) :
 \be\label{AVD}
 \lan D(\bm{x},\bm{y})\ran_\tau \,\equiv \,\int D[\rho]\,
 D(\bm{x},\bm{y})\,Z_{\tau}[\rho]\,\approx\,
 \frac{1}{2}\big(n_\tau(\bm{x},\bm{y}) +
 n_\tau(\bm{y},\bm{x})\big).
 \ee
\comment{
 and therefore (compare to Eq.~(\ref{D0DM})) :
 \be\hspace*{-0.3cm}
 D(\bm{x},\bm{y})\,Z_{\tau}[\rho]\, &\,
 \approx\,&
 \sum_{N=1}^{\infty} \, \int d\Gamma_N \, P_N(\tau)
    \sum_{i=1}^N
\frac{1}{2}
 \big(\delta_{\bm{z}_{i-1}\bm{x}}\delta_{\bm{z}_{i}\bm{y}} +
 \delta_{\bm{z}_{i-1}\bm{y}}\delta_{\bm{z}_{i}\bm{x}}\big)
 \prod_{j}
     D^{\dagger}_0(\bm{z}_{j-1},\bm{z}_j) \,
     \delta[\rho],\nn \label{DZ}
 \ee
which immediately implies that $\lan D(\bm{x},\bm{y})\ran_\tau=
 (n_\tau(\bm{x},\bm{y}) + n_\tau(\bm{y},\bm{x}))/2$.}
More generally, when acting on $Z_{\tau}[\rho]$ with a string of
$k$ $D$--operators corresponding to non--identical dipoles, then
one measures the $k$--body dipole density (symmetrized under the
exchange of the quark and antiquark legs of each dipole); e.g.,
for $k=2$, one finds the analog of Eq.~(\ref{AVD0}). Note that,
although two operators like $D(\bm{x_1},\bm{y_1})$ and
$D(\bm{x_2},\bm{y_2})$ do not commute with each other, this
non--commutativity is irrelevant when computing their action on
the onium weight function and for large $N_c$.

In the remaining part of this section, we shall verify that the
above interpretation of the operator $D(\bm{x},\bm{y})$,
Eq.~(\ref{DA}), as the dipole number operator is consistent with
the evolution of the correlations of $\rho$ generated by the
Bremsstrahlung Hamiltonian (\ref{HBREM}) at large $N_c$. In Sect.
2, we have already performed this check for the 2--point function:
when acting on $\rho^a(\bm x) \rho^a(\bm y)$, $H_{\rm BREM}$
generates the BFKL equation (\ref{evolnumber}) for $\lan
D(\bm{x},\bm{y})\ran_\tau$, in agreement with Eq.~(\ref{AVD}). In
what follows, we shall use the results of Sect. 3 to perform the
corresponding check for the dipole pair density. That is, we shall
verify that the evolution equation satisfied at large $N_c$ by the
following 4--point correlation function:
\begin{align}
\langle
D(\bm{x}_1,\bm{y}_1)D(\bm{x}_2,\bm{y}_2)\rangle_\tau=\frac{1}{g^4N_c^2}\langle\rho^a(\bm{x}_1)
\rho^a(\bm{y}_1) \rho^b(\bm{x}_2)\rho^b(\bm{y}_2)\rangle_\tau,
\end{align}
is consistent with the known evolution equation for
$n^{(2)}(\bm{x}_1,\bm{y}_1;\bm{x}_2,\bm{y}_2)$, including the
interesting, `fluctuation', term (i.e., the term linear in $n$
which describes the formation of the dipole pair
$\{(\bm{x}_1,\bm{y}_1),\,(\bm{x}_2,\bm{y}_2)\}$ through the
splitting of one dipole in the last step of the evolution)
\cite{LL04,IT04}.


Using Eq.~(\ref{eq:BalNc2}), we obtain
\begin{align}\frac{\partial}{\partial \tau}\langle D(\bm{x}_1,\bm{y}_1)D(\bm{x}_2,\bm{y}_2)
 \rangle_\tau
=\frac{\bar{\alpha}_s}{2\pi}\sum_{N=1}^{\infty} \, \int d\Gamma_N
\, P_N(\tau) \sum_{i=1}^N \int\limits_{\bm{z}}
    \mathcal{M}_{\bm{u}_i\bm{v}_i\bm{z}}\int  D[\rho]\,
D(\bm{x}_1,\bm{y}_1)D(\bm{x}_2,\bm{y}_2)
   \nonumber \\  \times \big[
    D^{\dagger}(\bm{u}_{i},\bm{z})\,D^{\dagger}(\bm{z},\bm{v}_{i})
    -D^{\dagger}(\bm{u}_{i},\bm{v}_{i})\big]   \prod_{j\neq i}
     D^{\dagger}(\bm{u}_{j},\bm{v}_{j}) \,
     \delta[\rho]\, . \label{new}
\end{align}
To evaluate the r.h.s. of this equation, we need to move the
operators $DD$ all the way to the right using the commutator
Eq.~(\ref{DADCDNC}) and then set $D^{\dagger}=1$. Clearly, the
following commutator vanishes
\begin{align} [\,D_{\x_1\y_1}, \,[\,D_{\x_2\y_2},\,
D^{\dagger}_{\bm{u}_k\bm{v}_k}\,]\,] \qquad ({\rm any}\ k),
\end{align} unless the two external dipoles completely overlap, an
uninteresting situation that we exclude. Moreover, terms like
\begin{align}
[\,D_{\x_1\y_1},D^{\dagger}_{\bm{u}_j\bm{v}_j}]
[\,D_{\x_2\y_2},D^{\dagger}_{\bm{u}_k\bm{v}_k}] \qquad (j,k \neq
i),
\end{align}
do not contribute either, because the corresponding coefficient
vanishes after setting $D^{\dagger}=1$. Therefore, we can replace
the operator part in Eq.~(\ref{new}) with
\begin{align}
D_{\bm{x}_1\bm{y}_1}D_{\bm{x}_2\bm{y}_2} (
    D^{\dagger}_{\bm{u}_{i}\bm{z}}D^{\dagger}_{\bm{z}\bm{v}_{i}}
    -D^{\dagger}_{\bm{u}_{i}\bm{v}_{i}})   \prod_{j\neq i}
     D^{\dagger}_{\bm{u}_{j}\bm{v}_{j}} \nonumber \\ \to
     \big[\,
     D_{\bm{x}_1\bm{y}_1},\,
     (D^{\dagger}_{\bm{u}_{i}\bm{z}}\,D^{\dagger}_{\bm{z}\bm{v}_{i}}
    -D^{\dagger}_{\bm{u}_{i}\bm{v}_{i}})\,\big]\big(\sum_{j\neq i}\,
    [\,D_{\bm{x}_2\bm{y}_2},
    \,D^{\dagger}_{\bm{u}_{j}\bm{v}_{j}}\,]\big)
   + \{1\leftrightarrow 2\}  \nonumber \\
   +[\,D_{\bm{x}_1\bm{y}_1},\,D^{\dagger}_{\bm{u}_{i}\bm{z}}\,]\,
   [\,D_{\bm{x}_2\bm{y}_2},\,D^{\dagger}_{\bm{z}\bm{v}_{i}}\,]
   +\{1\leftrightarrow 2\}. \label{int}
\end{align}
The second line of Eq.~(\ref{int}) describes the BFKL evolution of
one of the two dipoles in the pair (and for large $N_c$). To see
this, note that, e.g.,
\begin{align}
   \sum_i \int\limits_{\bm{z}}
    \mathcal{M}_{\bm{u}_i\bm{v}_i\bm{z}}
    \big[\,D_{\bm{x}_1\bm{y}_1},\,
    (D^{\dagger}_{\bm{u}_{i}\bm{z}}\,D^{\dagger}_{\bm{z}\bm{v}_{i}}
    -D^{\dagger}_{\bm{u}_{i}\bm{v}_{i}})\,\big]
    \big\arrowvert_{D^{\dagger}=1}\nonumber \\ =
    \frac{1}{2}\sum_i\Big\{
    \mathcal{M}_{\bm{x}_1\bm{v}_i\bm{y}_1}\delta_{\x_1\bm{u}_i}+
    \mathcal{M}_{\bm{u}_i\bm{y}_1\bm{x}_1}\delta_{\y_1\bm{v}_i}
-
\int\limits_{\bm{z}}\mathcal{M}_{\bm{x}_1\bm{y}_1\bm{z}}\delta_{\x_1\bm{u}_i}\delta_{\y_1\bm{v}_i}
+ \{\x_1 \leftrightarrow \y_1\} \Big\} \nonumber \\
=\frac{1}{2}\sum_i \int\limits_{\bm{z}} \Big\{
  \mathcal{M}_{\bm{x}_1\bm{z}\bm{y}_1}\delta_{\x_1\bm{u}_i}\delta_{\bm{z}\bm{v}_i}+
   \mathcal{M}_{\bm{z}\bm{y}_1\bm{x}_1}\delta_{\bm{z}\bm{u}_i}\delta_{\y_1\bm{v}_i}
 -\mathcal{M}_{\bm{x}_1\bm{y}_1\bm{z}}\delta_{\x_1\bm{u}_i}\delta_{\y_1\bm{v}_i}
+ \{\x_1 \leftrightarrow \y_1\} \Big\} ,
\end{align}
which represents the BFKL evolution of the dipole
$(\bm{x}_1,\bm{y}_1)$, cf. Eq.~(\ref{evolnumber}).

The last line of Eq.~(\ref{int}) is the fluctuation term that we
are primarily interested in. For large $N_c$, the commutators
there can be evaluated according to Eq.~(\ref{DADCDNC}). Note that
the use of Eq.~(\ref{DADCDNC}) (instead of the exact relation
(\ref{DDCOM})) automatically avoids contributions in which the
external dipoles get mixed with each other under the action of the
commutators (that is, contributions where the two factors of
$\rho$ from a same external dipole get contracted with factors of
$W$ coming from different internal dipoles); this was to be
expected, since such contributions are indeed suppressed at large
$N_c$.

After also integrating over $\z$ and summing over $N$, we finally
obtain
\begin{align} \frac{\partial}{\partial
\tau}\langle  D_{\bm{x}_1,\bm{y}_1}D_{\bm{x}_2,\bm{y}_2}
\rangle_\tau &=
 \big[ H_{\rm BFKL}^{(1)}+H_{\rm BFKL}^{(2)} \big]\langle
 D_{\bm{x}_1,\bm{y}_1}D_{\bm{x}_2,\bm{y}_2}\rangle_{\tau}
\nonumber \\ &+\frac{\bar{\alpha}_s}{4\pi}\Bigl\{
\mathcal{M}_{\bm{x}_1\bm{y}_2\bm{x}_2}\delta_{\bm{x}_2\bm{y}_1}\langle
D_{\bm{x}_1\bm{y}_2} \rangle_\tau
+\mathcal{M}_{\bm{x}_1\bm{x}_2\bm{y}_1}\delta_{\bm{y}_1\bm{y}_2}
\langle D_{\bm{x}_1\bm{x}_2} \rangle_\tau \nonumber
\\ &+ \ \mathcal{M}_{\bm{y}_1\bm{y}_2\bm{x}_1}\delta_{\bm{x}_1\bm{x}_2}\langle
D_{\bm{y}_1\bm{y}_2} \rangle_\tau
+\mathcal{M}_{\bm{y}_1\bm{x}_2\bm{x}_1}\delta_{\bm{x}_1\bm{y}_2}
\langle D_{\bm{y}_1\bm{x}_2} \rangle_\tau \Bigr\}, \label{ab}
\end{align}
with the compact notation
 \begin{align}
 H_{\rm BFKL}^{(1)} \langle
 D_{\bm{x}_1,\bm{y}_1}D_{\bm{x}_2,\bm{y}_2}\rangle_{\tau}\,\equiv
 \ & \frac{\abar}{2\pi} \int_{\bm{z}}\,
    \big[
    -\, {\cal M} ({\bm{x}_1},{\bm{y}_1},{\bm{z}})\,\langle
 D_{\bm{x}_1,\bm{y}_1}D_{\bm{x}_2,\bm{y}_2}\rangle_{\tau}
    \nonumber \\
    +&\, {\cal M} ({\bm{x}_1},{\bm{z}},{\bm{y}_1})\,\langle
 D_{\bm{x}_1,\bm{z}}D_{\bm{x}_2,\bm{y}_2}\rangle_{\tau}
    + {\cal M} ({\bm{z}},{\bm{y}_1},{\bm{x}_1})\,\langle
 D_{\bm{z},\bm{y}_1}D_{\bm{x}_2,\bm{y}_2}\rangle_{\tau}
    \big].
\end{align}
In identifying the expectation values in the r.h.s.'s of the above
equations, we have used the (dressed--dipole version of the)
relations (\ref{AVD0}) and (\ref{AVD02}). The fluctuation terms
are the terms proportional to the average dipole number density
$\langle D \rangle_\tau$ in Eq.~(\ref{ab}). Using Eq.~(\ref{AVD0})
and (\ref{AVD02}) once again, one can check that Eq.~(\ref{ab}) is
equivalent to Eq.~(5.15) of Ref.~\cite{IT04}. This confirms the
interpretation of $\langle DD \rangle_{\tau}$ as the average
dipole pair density.

\section{Bremsstrahlung effects on dipole--onium scattering}
\setcounter{equation}{0}

So far, we have almost exclusively focused on the evolution of the
target wavefunction : the large--$N_c$ evolution in the dilute
regime has been formulated either as an evolution equation for the
onium weight function,  Eq.~(\ref{eq:BalNc2}), or as a set of
coupled evolution equations for those correlations of $\rho$ which
have the meaning of dipole densities (see, e.g.,
Eqs.~(\ref{evolnumber}) and (\ref{ab})). But the scattering
amplitudes for external projectiles involve also different
correlations of $\rho$ (or, more precisely, of $\alpha\equiv
\alpha_\infty^a(\x)$; see, e.g., Eq.~(\ref{AVTdipole})), so it
would be interesting to establish the corresponding evolution
equations as well, at least at large $N_c$. As anticipated in the
Introduction, the evolution equations for scattering amplitudes
generated by $H_{\rm BREM}$ are {\em a priori} more general than
those previously derived in the literature from the dipole picture
\cite{IT04,MSW05,IT05}, in the sense of including additional
processes in which individual target dipoles exchange more than
two gluons with the projectile. However, our main purpose in this
section is not to derive such more general equations, but rather
to give an argument, based on an explicit example, that the
additional effects are in fact suppressed at sufficiently high
energy.

The simplest scattering problem which is sensitive to
Bremsstrahlung (i.e., to gluon number fluctuations) in the target
wavefunction is the scattering with two external dipoles. (For a
single external dipole, $H_{\rm BREM}$ generates the standard BFKL
equation; see below.) To lowest order in perturbation theory, each
external dipole can exchange two gluons with the target, cf.
Eq.~(\ref{Tdipole}). Therefore, the leading--order contribution to
the amplitude for the {\em simultaneous} scattering of the
incoming dipoles reads
 \be\label{T2dipole_weak} \langle T^{(2)}(\x_1,\y_1;\x_2,\y_2)
\rangle_\tau \, \simeq \,
  \frac{g^4}{16N_c^2}\,\big \langle
 (\alpha_a(\x_1)-\alpha_a(\y_1))^2\, (\alpha_b(\x_2)-\alpha_b(\y_2))^2
 \big\rangle_\tau\,,\ee
and involves a total exchange of four gluons (two with each
external dipole). These four gluons can be absorbed either by two
different dipoles in the target wavefunction, or by a single such
a dipole, and the two type of processes are parametrically of the
same order in $\alpha_s$ and $1/N_c$.  In particular, they both
contribute in the large--$N_c$ limit. Still, as we shall argue
below, the corresponding contributions behave differently when
increasing the energy, in such a way that multiple exchanges with
the same target dipole are relatively suppressed at high energy.

To see this, we need to evaluate the average in
Eq.~(\ref{T2dipole_weak}) with the onium weight function
(\ref{ZDipole}). The scattering operator for a single (external)
dipole, that is,
  \be\label{T0dipole} T_0(\x,\y) &\equiv &
  \frac{g^2}{4N_c}\,
 \big(\alpha^a({\x})-\alpha^a({\y})\big)^2\,,\ee
can be conveniently expressed in terms of the (target) dipole
number operator introduced in Sect. 4. Namely, after inverting the
Poisson equation (\ref{Poisson}) to relate $\alpha^a$ to $\rho^a$,
one obtains
\begin{align}\label{calG}
  \alpha^a({\x})-\alpha^a({\y})= \int\limits_{\bm{u}}
 \cal{G}(\bm{u}|\bm{x},\bm{y})\,\rho^a(\bm{u})\,,\qquad
    \cal{G}(\bm{u}|\bm{x},\bm{y}) \equiv
    \frac{1}{4\pi}
    \ln \frac{(\bm{u}-\bm{y})^2}{(\bm{u}-\bm{x})^2}\,,
\end{align}
which then allows us to successively write
 \be\label{TD}
 T_0(\x,\y)&=&\,\frac{g^2}{4N_c}\,\int\limits_{\bm{u},\bm{v}}\,
  \cal{G}(\bm{u}|\bm{x},\bm{y})
  \cal{G}(\bm{v}|\bm{x},\bm{y})\,\rho^a(\bm{u})\rho^a(\bm{v})\nn
 &{}& =\,-\,\frac{g^2}{8N_c}\,\int\limits_{\bm{u},\bm{v}}\,
 \big[\cal{G}(\bm{u}|\bm{x},\bm{y}) -
  \cal{G}(\bm{v}|\bm{x},\bm{y})\big]^2\,\rho^a(\bm{u})\rho^a(\bm{v})\nn
 &{}& =\, \int\limits_{\bm{u},\bm{v}}\,
 \cal{A}_0(\bm{x},\bm{y}|\bm{u},\bm{v})\, D(\bm{u},\bm{v}),
  \ee
where in going from the first to the second line we have used the
fact that the system is globally color neutral : $\int_{{\bm u}}
\rho^a(\bm{u}) =0$. The operator $D(\bm{u},\bm{v})$ has been
defined in Eq.~(\ref{DA}), and $\cal{A}_0$ is the amplitude for
dipole--dipole scattering in the two--gluon exchange approximation
and for large $N_c$ :
\begin{align}\label{T0}
    \cal{A}_0(\bm{x},\bm{y}|\bm{u},\bm{v}) =
    \frac{\alpha_s^2}{8}
    \left[\ln \frac{(\bm{x}-\bm{v})^2 (\bm{y}-\bm{u})^2}
    {(\bm{x}-\bm{u})^2 (\bm{y}-\bm{v})^2}
    \right]^2.
\end{align}
It is now straightforward to compute the average scattering
amplitude for a single external dipole: using Eq.~(\ref{DZN})
together with the symmetry of the dipole--dipole amplitude
$\cal{A}_0(\bm{x},\bm{y}|\bm{u},\bm{v})$ under the exchange
$\bm{u} \leftrightarrow \bm{v}$, one immediately obtains
 \be\label{TAV} \langle T(\x,\y)\rangle_\tau
 \,=\, \int\limits_{\bm{u},\bm{v}}\,
 \cal{A}_0(\bm{x},\bm{y}|\bm{u},\bm{v})\, n_\tau(\bm{u},\bm{v}),
  \ee
which is the relation  expected within the dipole picture
\cite{AM94,AM95,IM031}. By  using this relation together with
Eq.~(\ref{evolnumber}) for $n_\tau$, one can show \cite{IM031}
that $ \langle T(\x,\y)\rangle_\tau$ obeys the standard BFKL
equation, as anticipated.

To similarly evaluate Eq.~(\ref{T2dipole_weak}), we shall use
Eq.~(\ref{TD}) for both external dipoles. We have
 \be\label{T2dipole1}\hspace*{-0.5cm}
  \langle T^{(2)}(\x_1,\y_1;\x_2,\y_2)
 \rangle_\tau \, = \, \int\limits_{\bm{u}_i,\bm{v}_i}
    \cal{A}_0(\bm{x}_1,\bm{y}_1|\bm{u}_1,\bm{v}_1)\,
    \cal{A}_0(\bm{x}_2,\bm{y}_2|\bm{u}_2,\bm{v}_2)\,
   \langle D(\bm{u}_1,\bm{v}_1)\,D(\bm{u}_2,\bm{v}_2)
   \rangle_\tau,\nn\ee
where after using again Eq.~(\ref{DZN}) and  the symmetries of the
tree--level amplitude $\cal{A}_0$, one can effectively replace (at
large $N_c$)
 \be\hspace*{-0.6cm}
\langle D(\bm{u}_1,\bm{v}_1)\,D(\bm{u}_2,\bm{v}_2)
   \rangle_\tau &\,\to\,& \big\langle
  \sum_{i=1}^N
    \delta^{(2)}(\bm{z}_{i-1}-\bm{u}_1)
    \delta^{(2)}(\bm{z}_i-\bm{v}_1)\,
    \sum_{k=1}^N
    \delta^{(2)}(\bm{z}_{k-1}-\bm{u}_2)
    \delta^{(2)}(\bm{z}_k-\bm{v}_2)\big\rangle_\tau\nn
    &{}&=\,
    n^{(2)}_\tau(\bm{u}_1,\bm{v}_1;\bm{u}_2,\bm{v}_2)
    + \delta^{(2)}(\bm{u}_1-\bm{u}_2)
     \delta^{(2)}(\bm{v}_1-\bm{v}_2)n_\tau(\bm{u}_1,\bm{v}_1).\nn
     \ee
We have recognized here the average dipole density in the target
$n_\tau$ and also the average dipole pair density $n^{(2)}_\tau$
according to their definitions in
Eqs.~(\ref{densityN})--(\ref{n2def1}). When the above expression
is inserted in the r.h.s. of Eq.~(\ref{T2dipole1}), we finally
obtain:
 \begin{align}\label{T2conv}
    \langle T^{(2)}(\x_1,\y_1;\x_2,\y_2)
   \rangle_\tau  =\,&
        \int\limits_{\bm{u}_i,\bm{v}_i}
    \cal{A}_0(\bm{x}_1,\bm{y}_1|\bm{u}_1,\bm{v}_1)\,
    \cal{A}_0(\bm{x}_2,\bm{y}_2|\bm{u}_2,\bm{v}_2)\,
    n^{(2)}_\tau(\bm{u}_1,\bm{v}_1;\bm{u}_2,\bm{v}_2)\nn
    & +
    \int\limits_{\bm{u},\bm{v}}\,
    \cal{A}_0(\bm{x}_1,\bm{y}_1|\bm{u},\bm{v})\,
    \cal{A}_0(\bm{x}_2,\bm{y}_2|\bm{u},\bm{v})\,
    n_\tau(\bm{u},\bm{v}),
\end{align}
with a clear physical interpretation for the two terms in the
r.h.s. : The first term, proportional to $n_\tau^{(2)}$, describes
the scattering with two {\em different} dipoles in the target,
while the second term, proportional to  $n_\tau$, represents a
four--gluon exchange with a {\em single} dipole. Incidentally, the
above derivation of Eq.~(\ref{T2conv}) confirms that no ordering
ambiguities appear in the evaluation of the scattering amplitudes
at large $N_c$.

Previous applications of the dipole picture to scattering
\cite{AM94,AM95,IM031,IT04,IT05} were all based on the assumption
that a dipole can exchange only two gluons, and thus they have
ignored the second term in Eq.~(\ref{T2conv}). Using only the
first term there, together with the known equation for
$n_\tau^{(2)}$ (essentially, Eq.~(\ref{ab})), one has derived an
evolution equation for $\langle T^{(2)}\rangle_\tau$
\cite{IT04,IT05} which includes the essential `fluctuation term'
through which $\langle T^{(2)}\rangle_\tau$ gets built from $
\langle T\rangle_\tau$ in the dilute regime. (This is induced by
the fluctuation terms in Eq.~(\ref{ab}).) In principle, one can
similarly use Eq.~(\ref{T2conv}) together with the known equations
for $n_\tau^{(2)}$ and $n_\tau$ to deduce the most general
evolution equation for $\langle T^{(2)}\rangle_\tau$ in the dilute
regime and at large $N_c$. In practice, this might be however
tedious, as it requires to invert Eqs.~(\ref{TAV}) and
(\ref{T2conv}) in order to express $n_\tau^{(2)}$ and $n_\tau$ in
terms of the scattering amplitudes.

But even without any detailed calculation, it is clear by
inspection of Eq.~(\ref{T2conv}) that at high energy (while still
in the dilute regime, though, for the dipole picture to apply),
the contribution involving the dipole pair density $n_\tau^{(2)}$
is in fact the dominant one: Indeed, with increasing energy,
$n_\tau$ grows like a BFKL pomeron, $n_\tau\sim
\exp\{\omega\bar\alpha_s\tau\} $, whereas $n_\tau^{(2)}$ grows
like a Pomeron squared: $n_\tau^{(2)}\sim
\exp\{2\omega\bar\alpha_s\tau\} $ ($\omega$ is a pure number).
Therefore, the contribution proportional to $n_\tau^{(2)}$
dominates as soon as $\bar\alpha_s\tau\simge 1$, which is the
interesting regime at high energy\footnote{This condition leaves a
parametrically large window for the applicability of the dipole
picture, which ceases to be valid when $n_\tau\sim 1/\alpha_s^2$
or $\bar\alpha_s\tau\sim \ln(1/\alpha_s^2)$ \cite{AM94,IT04}.}.
This implies not only that, for $\bar\alpha_s\tau\simge 1$, one
can neglect $n_\tau$ next to $n_\tau^{(2)}$ in the r.h.s. of
Eq.~(\ref{T2conv}), but also that, in writing an equation for
$\langle T^{(2)}\rangle_\tau$ which should describe its evolution
from the low energy regime at $\tau\sim 0$ up to the high energy
where $\bar\alpha_s\tau\simge 1$, it is enough to keep trace of
the terms coming from the evolution of $n_\tau^{(2)}$. This is
what has been done in Ref. \cite{IT04,IT05}.

Although certainly correct (in view of the previous arguments),
the last statement is nevertheless quite subtle, as it can be
appreciated when trying to identify the fluctuation term (the
contribution proportional to $ \langle T\rangle_\tau$) in the
evolution equation for $\langle T^{(2)}\rangle_\tau$ derived from
Eq.~(\ref{T2conv}). By taking a derivative w.r.t. $\tau$ in
Eq.~(\ref{T2conv}) and using the equations for $n_\tau^{(2)}$ and
$n_\tau$, one finds two type of contributions which are
proportional to $n_\tau$ (and thus to $ \langle T\rangle_\tau$,
cf. Eq.~(\ref{TAV})) : the `genuine' fluctuation term coming from
the evolution (\ref{ab}) of $n_\tau^{(2)}$, and the terms
describing the BFKL evolution (\ref{evolnumber}) of $n_\tau$. Both
types of contributions are parametrically of the same order ---
namely, of order $\bar\alpha_s\alpha_s^2\langle T\rangle_\tau$---,
and thus contribute on equal footing to the growth of $\langle
T^{(2)}\rangle_\tau$ in the dilute regime. This seems to
contradict the results in Ref. \cite{IT04,IT05}, where only the
`genuine' fluctuation term has been included.

However, this contradiction is only illusory: First, to properly
identify the fluctuation terms, one need to reexpress everywhere
$n_\tau^{(2)}$ in terms of $\langle T^{(2)}\rangle_\tau$,
according to Eq.~(\ref{T2conv}); this operation modifies the
fluctuation terms since, schematically, $\cal{A}_0^2\,n_\tau^{(2)}
= \langle T^{(2)}\rangle_\tau - \cal{A}_0^2\, n_\tau$. Second, the
additional fluctuation terms (besides the `genuine' one) which
persists after the previous operation have only the role to
compensate the spurious double--Pomeron contribution which emerges
from the evolution of that piece of the initial condition $\langle
T^{(2)}\rangle_{0}$ which is introduced by the second term in
Eq.~(\ref{T2conv}). Therefore, if one starts with the following
initial conditions at $\tau=0$ (i.e., a target made with a single
dipole):
 \be\label{initial}
 n_0(\bm{u},\bm{v})=\delta^{(2)}(\bm{u}-\bm{u}_0)
  \delta^{(2)}(\bm{v}-\bm{v}_0),\qquad n_0^{(2)} = 0\,,\ee
which in turn implies
\begin{align}\label{T20}
    \langle T^{(2)}(\x_1,\y_1;\x_2,\y_2)
   \rangle_0  =
    \cal{A}_0(\bm{x}_1,\bm{y}_1|\bm{u}_0,\bm{v}_0)\,
    \cal{A}_0(\bm{x}_2,\bm{y}_2|\bm{u}_0,\bm{v}_0)\,
    ,
\end{align}
then the dominant contribution to $\langle T^{(2)}\rangle_\tau$ at
high energy --- the one which grows like a double--Pomeron, $\sim
\exp\{2\omega\bar\alpha_s\tau\} $
--- comes entirely from the `genuine' fluctuation term, and {\em
not} from the evolution of the above initial condition. The latter
will give only a subleading contribution, $\sim
\exp\{\omega\bar\alpha_s\tau\} $.

These considerations can be clarified with the help of a simple
example, borrowed from Ref. \cite{IT04}, which has the advantage
to be easily solvable while keeping the non--trivial features of
interest here\footnote{It is likely that the following argument is
similar to that developed in a different context by Braun and
Vacca \cite{BV99}, but we have not been able to clearly establish
the correspondence between the two problems. See also the
discussion in Ref. \cite{IT05}.}. Namely, let us replace the
dipoles par point--like particles which live at a fixed point (so
there is no spatial dimension involved in their dynamics), and
whose number distribution evolves according to the following
hierarchy of equations:
 \be\label{toylinevoln}
    \frac{d n_\tau}{d\tau}\,=\,
    \alpha \,n_\tau,\qquad
    \frac{d n_\tau^{(2)}}{d\tau}\,=\,
    2\alpha \left[
    n_\tau^{(2)} + \, n_\tau
    \right],\quad\dots
 \ee
which mimic the equations satisfied by the dipole densities in the
dipole picture. The solution corresponding to the initial
conditions $n(0)=n_0$ and $n^{(2)}(0)=0$ reads:
\begin{align}\label{toysol1}
    n_\tau &= n_0 \exp(\alpha \tau),
    \nn 
    n^{(2)}_\tau &=
     2 n_0
    \,\exp(2 \alpha \tau)
    -2 n_0 \exp(\alpha \tau),\,\,\dots
\end{align}
where $n^{(2)}_\tau$ has been generated  by the term linear in
$n_\tau$ in the r.h.s. of the second equation (\ref{toylinevoln});
this is the analog of the `genuine fluctuation term' in the
present model, and plays the role of a source for $n^{(2)}_\tau$
(so like the actual fluctuation terms in Eq.~(\ref{ab})). If one
further introduces the analog of the `scattering amplitudes' :
 \be\label{T2toy0}
  T_\tau = n_\tau, \qquad T^{(2)}_\tau = A n^{(2)}_\tau + B n_\tau,
 \ee
then clearly
 \be\label{T2toy}
  T^{(2)}_\tau = 2An_0
    \,[\exp(2 \alpha \tau)
    -\exp(\alpha \tau)] + Bn_0 \exp(\alpha \tau)\,\approx\,
  2An_0\,\exp(2 \alpha \tau),\ee
where the approximate equality holds at large time ($\alpha
\tau\gg 1$), and the corresponding contribution to $T^{(2)}_\tau$
is entirely coming from $n^{(2)}_\tau$ (like in
Eq.~(\ref{T2conv})). Let us now construct the evolution equation
for  $T^{(2)}_\tau$ and follow the fluctuation terms:
 \be\label{T2eq1}
  \frac{d T_\tau^{(2)}}{d\tau}&\,=\,&A
 \frac{d n_\tau^{(2)}}{d\tau}\,+\,B
 \frac{d n_\tau}{d\tau}\,=\,
    2A\alpha \left[
    n_\tau^{(2)} + \, n_\tau
    \right] + B\alpha \,n_\tau\nn
  &\,=\,& 2\alpha \,T_\tau^{(2)} + (2A-B)\alpha\,T_\tau,\ee
where the second line identifies $(2A-B)\alpha\,T_\tau$ as the
fluctuation term. This involves the two types of contributions
alluded to before: $2A \alpha\,T_\tau$ is the genuine fluctuation
term introduced by the evolution of $n^{(2)}_\tau$, whereas
$(-B)\alpha\,T_\tau$ has been induced by the BFKL evolution of
$n_\tau$. With $T_\tau = n_0 \exp(\alpha \tau)$, the above
equation is solved by
 \be\label{T2toy1}
  T^{(2)}_\tau &= &T^{(2)}_0\exp(2 \alpha \tau)
  + (2A-B)n_0
    \,[\exp(2 \alpha \tau)
    -\exp(\alpha \tau)]
     ,\ee
where both types of fluctuations seem to contribute to the
dominant, `double--Pomeron', rise at large time. However, by
recalling that $T^{(2)}_0 = Bn_0$, cf. Eq.~(\ref{T2toy0}), one
immediately sees that the `double--Pomeron' terms proportional to
$B$ do actually cancel between the contribution of the initial
condition and that of the fluctuation terms. The remaining
`double--Pomeron' term,  proportional to $A$, is the one generated
by the genuine fluctuation term, so like in the direct calculation
leading to Eq.~(\ref{T2toy}). Thus, the same dominant behaviour at
large time would have been obtained by solving the simplified
equation:
 \be\label{T2eq2}
  \frac{d T_\tau^{(2)}}{d\tau}\,=\,2\alpha
   \, T_\tau^{(2)} + 2A\alpha\, T_\tau
   ,\ee
(which arises by assuming that $T^{(2)}_\tau = A  n^{(2)}_\tau$)
together with the initial condition $T^{(2)}_0 =0$.
Eq.~(\ref{T2eq2}) is the analog of the equation for $\langle
T^{(2)}\rangle_\tau$ derived in Refs. \cite{IT04,IT05}, whereas
Eq.~(\ref{T2eq1}) corresponds to the more complete equation that
would be obtained in QCD at large $N_c$ after including the
effects of Bremsstrahlung in the target wavefunction.

The simple example above also emphasizes the importance of
correctly adjusting the initial condition to the approximations
that we perform on the evolution equation: If, as in Refs.
\cite{IT04,IT05}, we restrict ourselves to the two--gluon exchange
approximation in the construction of the evolution equation for
$\langle T^{(2)}\rangle_\tau$, then the same approximation must be
performed also on the initial condition. For instance, if at
$\tau=0$ the target is a bare dipole, cf. Eq.~(\ref{initial}),
then the initial condition to be used for the equations in Refs.
\cite{IT05} is $\langle T^{(2)}\rangle_0=0$, and not
Eq.~(\ref{T20}). This amounts to consistently neglect
contributions like the second term in Eq.~(\ref{T2conv}) at all
places.

\section{Conclusions}
\setcounter{equation}{0}

In this paper we have shown that the recently developed theory for
Bremsstrahlung in the QCD evolution with increasing energy
\cite{KL05,KL3,BREM} is consistent with the dipole picture in the
large--$N_c$ limit. The characteristic feature of this theory is
that the quark and the antiquark parts of a color dipole are
represented as color sources which can radiate arbitrarily many
gluons with relatively small longitudinal momenta. This
generalizes previous `color glass' descriptions of the onium
wavefunction, in which the individual dipoles were allowed to
radiate only two gluons \cite{IM031,IT04,MSW05,IT05}. The energy
evolution of the ensemble of color sources is known for arbitrary
$N_c$ --- this is described by a two--dimensional, Hamiltonian,
field theory with SU$(N_c)$--like Poisson brackets
---, and our objective in this paper has been to
demonstrate that, for large $N_c$, this evolution can be
reformulated in terms of color dipoles which evolve through dipole
splitting. To that aim, we have identified the operator which
describes, within the Hamiltonian theory, a quark--antiquark
dipole `dressed' by the radiation, and then we have shown that, at
large $N_c$, the action of the Hamiltonian on a collection of such
dipoles consists in the splitting of any of the original dipoles
into two new dipoles with one common leg. This confirms the fact
that the effective degrees of freedom for high energy evolution in
QCD at large $N_c$ and in the dilute regime are quark--antiquark
color dipoles, as originally demonstrated at an abstract level
(i.e., without specifying the way how the dipoles are actually
measured) in the pioneering papers by Mueller \cite{AM94,AM95}.

Whereas the emergence of the dipole picture at large $N_c$ was to
be expected (in view of the general results in Ref. \cite{AM94}),
the technical manipulations necessary to demonstrate it turned out
to be quite complex. In particular, we have found that the {\em
structure} of the Hamiltonian cannot be further simplified when
going to the large--$N_c$ limit. Rather, it is its {\em action} on
the `onium weight function' (built with specific dipole creation
operators) which reduces to dipole splitting at large $N_c$. Thus
the present construction provides a rather subtle generalization
of previous formalisms based on the two--gluon exchange
approximation \cite{IM031,MSW05}: Whereas the structure of the
wavefunction in terms of dipole creation operators is formally the
same (with different meanings for the creation operators though),
the corresponding Hamiltonians are very different, and cannot be
directly related to each other via some large--$N_c$
approximations.

This situation is reminiscent of that encountered in a recent
large--$N_c$ analysis of the JIMWLK evolution in the high density
regime \cite{BIIT05}, and in fact the duality between the two
Hamiltonian field theories (JIMWLK and Bremsstrahlung)
\cite{KL3,BREM} has played an important role in the present
analysis. Namely, the action of the JIMWLK Hamiltonian on
scattering operators for external dipoles turns out to be dual to
that of the Bremsstrahlung Hamiltonian on creation operators for
internal dipoles. Because of that, many of the technical
manipulations in this paper are similar (more properly, dual) to
those encountered in the derivation of the Balitsky equations
\cite{B} from the JIMWLK equation \cite{W,RGE}.

We have finally investigated the effects of Bremsstrahlung on the
scattering between two external dipoles and a dilute onium. We
have shown that, although some new effects appear, as associated
with the double scattering off a same target dipole, such effects
are in fact suppressed at high energy as compared to those already
included in the two--gluon exchange approximation. This
conclusion, which is corroborated by the recent analysis in Ref.
\cite{MMSW05}, implies that the evolution equations with `pomeron
loops' derived in Refs. \cite{IT04,MSW05,IT05} are indeed the
correct equations in QCD at large $N_c$ and for sufficiently high
energy.

It remains as an interesting open problem to consider the
generalization of some of the results obtained here (in
particular, in relation with the scattering problem) to arbitrary
$N_c$. Although the corresponding evolution Hamiltonian is known,
and so is also its action in terms of Poisson brackets (cf. Sect
2), we expect the corresponding analysis to be complicated by the
issue of the ordering of the operators. This is a new type of
problem, which had not been encountered before in the framework of
the color glass formalism, and so far it is not even clear whether
this formalism can be extended to account for the
non--commutativity of the color charges (see, e.g., the
discussions in Refs. \cite{KL05,KL3,KL4,BREM,MMSW05}). One may
however expect to be able to establish more explicit connections
to previous approaches within perturbative QCD, which are aiming
at the direct calculation of the $2\to n$ gluon vertices at high
energy and for arbitrary $N_c$ \cite{BW95,BV99,BE99,BLV05}.

\section*{Acknowledgments}

This work has been initiated in a collaboration with Dionysis
Triantafyllopoulos to whom we would like to thank for his help in
the early stages and many subsequent discussions. We are grateful
to Al Mueller for insightful conversations and for having informed
us about his recent work in collaboration with Cyrille Marquet,
Arif Shoshi and Stephen Wong, which leads to conclusions similar
to ours. One of us (E.I.) acknowledges useful conversations with
Jean--Paul Blaizot, Kazu Itakura, and Cyrille Marquet. Y. H. is
supported by Special Postdoctoral Research Program of RIKEN.  This
research has been partially supported by the Polish Committee for
Scientific Research, KBN Grant No. 1 P03B 028 28. This manuscript
has been authorized under Contract No. DE-AC02-98CH10886 with the
U. S. Department of Energy.


\begin{thebibliography}{10}

\bibitem{AM94}
A.H.~Mueller, {\it Nucl. Phys.} {\bf B415} (1994) 373; A.H.
Mueller and B.  Patel, {\it Nucl. Phys.} {\bf B425} (1994) 471.

\bibitem{AM95}
A.H.~Mueller, {\it Nucl. Phys.} {\bf B437} (1995) 107.

\bibitem{IM031}
E.~Iancu and A.H.~Mueller, {\it Nucl.\ Phys.}\ {\bf A730} (2004)
460.

\bibitem{IT04}
E.~Iancu and D.N.~Triantafyllopoulos, {\it ``A Langevin equation
for high energy evolution with Pomeron Loops''},
arXiv:hep-ph/0411405 [to appear in
  {\it Nucl.~Phys.~}{\bf A}].

\bibitem{MSW05}
A.H.~Mueller, A.I.~Shoshi and S.M.H.~Wong, {\it Nucl.\ Phys.}\
{\bf B715} (2005) 440.

\bibitem{IT05}
E.~Iancu and D.N.~Triantafyllopoulos, {\it Phys.~Lett.~}{\bf B610}
(2005) 253.

\bibitem{BIIT05}
J.-P.~Blaizot, E.~Iancu, K.~Itakura and D.N.~Triantafyllopoulos,
{\it Phys.~Lett.~}{\bf B615} (2005) 221.

\bibitem{KL05}
A.~Kovner and M.~Lublinsky, {\it Phys.~Rev.~}{\bf D71} (2005)
085004.

\bibitem{KL3}
A.~Kovner and M.~Lublinsky, {\it Phys. Rev. Lett.} {\bf 94} (2005)
181603. 

\bibitem{BREM}
Y.~Hatta, E.~Iancu, L.~McLerran, A.M.~Stasto, and
D.N.~Triantafyllopoulos,   {\it
  ``Effective Hamiltonian for QCD evolution at high energy''},
  arXiv:hep-ph/0504182.

\bibitem{KL4}
A.~Kovner and M.~Lublinsky, {\it ``Dense-Dilute Duality at work:
dipoles of the
  target''}, arXiv:hep-ph/0503155.

\bibitem{MV}
L.~McLerran and R.~Venugopalan, {\it Phys.\ Rev.}\ {\bf D49}
(1994) 2233; {\it
  ibid.} {\bf 49} (1994) 3352; {\it ibid.} {\bf 50} (1994) 2225.

\bibitem{RGE}
E.~Iancu, A.~Leonidov and L.~McLerran, {\it Nucl. Phys.}~{\bf
A692} (2001) 583;
  {\it Phys. Lett.} {\bf B510} (2001) 133; E.~Ferreiro, E.~Iancu, A.~Leonidov
  and L.~McLerran, {\it Nucl. Phys.} {\bf A703} (2002) 489.

\bibitem{CGCreviews}
  E.~Iancu, A.~Leonidov and L.~McLerran, {\it ``The Colour Glass Condensate: An
  Introduction''}, arXiv:hep-ph/0202270. Published in {\it QCD Perspectives on
  Hot and Dense Matter}, Eds. J.-P.~Blaizot and E.~Iancu, NATO Science Series,
  Kluwer, 2002;\\ E.~Iancu and R.~Venugopalan, {\it ``The Color Glass
  Condensate and High Energy Scattering in QCD''}, arXiv:hep-ph/0303204.
  Published in {\it Quark-Gluon Plasma 3}, Eds. R.C.~Hwa and X.-N.~Wang, World
  Scientific, 2003;\\ H.~Weigert, {\it ``Evolution at small $x_{\rm bj}$: The
  Color Glass Condensate''}, arXiv:hep-ph/0501087.

\bibitem{LL03}
E.~Levin and M.~Lublinsky, {\it Nucl.\ Phys.}\ {\bf A730} (2004)
191.

\bibitem{LL04}
E.~Levin and M.~Lublinsky, {\it Phys. Lett.} {\bf B607} (2005)
131.

\bibitem{AMprivate}
A.H. Mueller, {\it private communication.}

\bibitem{MMSW05}
C. Marquet, A.H.~Mueller, A.I.~Shoshi and S.M.H.~Wong, {\it in
preparation}.


\bibitem{JKLW97}
J.~Jalilian-Marian, A.~Kovner, A.~Leonidov and H.~Weigert, {\it
Nucl.\ Phys.}\
  {\bf B504} (1997) 415; {\it Phys.\ Rev.}\ {\bf D59} (1999)
  014014;  J.~Jalilian-Marian, A.~Kovner and  H.~Weigert,
        {\it  Phys.\ Rev.}\ {\bf D59} (1999) 014015;
    A. Kovner, J. G. Milhano and H. Weigert,
    {\it Phys. Rev.} {\bf D62} (2000) 114005.

\bibitem{W}
H.~Weigert, {\it Nucl. Phys.} {\bf A703} (2002) 823.

\bibitem{ODDERON}
Y.~Hatta, E.~Iancu, K.~Itakura and L.~McLerran, {\it ``Odderon in
the Color
  Glass Condensate''}, arXiv:hep-ph/0501171  [to appear in
  {\it Nucl.~Phys.~}{\bf A}].

\bibitem{B}
I.~Balitsky, {\it Nucl.\ Phys.}\ {\bf B463} (1996) 99; {\it Phys.
Lett.} {\bf
  B518} (2001) 235; {\it ``High-energy QCD and Wilson lines''},
  arXiv:hep-ph/0101042.

\bibitem{JP04}
R.A. Janik, R.~Peschanski, {\it Phys. Rev.} {\bf D70}
    (2004) 094005;
    R.A.~Janik, {\it Phys. Lett.} {\bf B604} (2004) 192.


\bibitem{BV99}
M.~Braun and G.P.~Vacca, {\it Eur.~Phys.~J.~}{\bf C6} (1999) 147.


\bibitem{BW95}
J.~Bartels and M. W\"usthoff, Z. Phys. C {\bf 66} (1995) 157.

\bibitem{BE99} J.~Bartels and C.~Ewerz,
JHEP {\bf 9909} (1999) 026; C.~Ewerz, Phys.\ Lett.\ B {\bf 472}
(2000) 135; {\it ibid.} {\bf 512} (2001) 239;  C.~Ewerz and V.
Schatz, Nucl.\ Phys.\ A {\bf 736} (2004) 371.

\bibitem{BLV05}J.~Bartels, L.N.~Lipatov and G.P.~Vacca,
Nucl. Phys. {\bf B706} (2005) 391;  J.~Bartels, M.~Braun and
G.~P.~Vacca, ``{\it Pomeron vertices in perturbative QCD in
diffractive scattering},'' hep-ph/0412218.



\end{thebibliography}
\end{document}